\documentclass[aps,prb,final,reprint,twocolumn,floatfix,showpacs,superscriptaddress,notitlepage]{revtex4-1}
\usepackage{graphicx,amssymb,soul,color,amsmath,bm}
\usepackage{epstopdf,hyperref}
\usepackage{braket,dsfont}
\usepackage[mathcal]{euscript}

\hypersetup{colorlinks=true,citecolor=blue,linkcolor=magenta}

\sethlcolor{yellow}
\allowdisplaybreaks

\begin{document}

\title{%
\textcolor{blue}{%
Photoinduced Hund excitons in the breakdown of a two-orbital Mott insulator
}%
}

\author{Juli\'an Rinc\'on}
\affiliation{Perimeter Institute for Theoretical Physics, Waterloo, Ontario N2L 2Y5, Canada}

\author{Elbio Dagotto}
\affiliation{Department of Physics and Astronomy, The University of Tennessee, Knoxville, Tennessee 37996, USA}
\affiliation{Materials Science and Technology Division, Oak Ridge National Laboratory, Oak Ridge, Tennessee 37831, USA}

\author{Adrian E.\ Feiguin}
\affiliation{Department of Physics, Northeastern University, Boston, Massachusetts 02115, USA}

\date{\today}

\begin{abstract}
We study the photoinduced breakdown of a two-orbital Mott insulator and resulting metallic state.~%
Using time-dependent density matrix renormalization group, we scrutinize the real-time dynamics of the half-filled two-orbital Hubbard model interacting with a resonant radiation field pulse.~%
The breakdown, caused by production of doublon-holon pairs, is enhanced by Hund's exchange, which dynamically activates large orbital fluctuations.~%
The melting of the Mott insulator is accompanied by a high to low spin transition with a concomitant reduction of antiferromagnetic spin fluctuations.~%
Most notably, the overall time response is driven by the photogeneration of excitons with orbital character that are stabilized by Hund's coupling.~%
These unconventional ``Hund excitons'' correspond to bound spin-singlet orbital-triplet doublon-holon pairs. We study exciton properties such as bandwidth, binding potential, and size within a semiclassical approach. The photometallic state results from a coexistence of Hund excitons and doublon-holon plasma.
\end{abstract}


\maketitle



\section{Introduction}
Control over the electronic properties of quantum matter using radiation is an attainable route to efficiently and systematically understand nonequilibrium phenomena. The interaction of correlated matter with light fields has already uncovered a wealth of exotic ``hidden'' phases~\cite{Orenstein12,Kampfrath13} including light-induced superconductivity~\cite{Fausti11}, dielectric breakdown~\cite{Miyano97,Wall11,Schiffrin13,Schultze13}, photometallic states in organic insulators~\cite{Okamoto07}, and photodisruption of charge, magnetic, and orbital order in complex oxides~\cite{Polli07,Ehrke11,Zhang16,Casals16}.

For dielectric breakdown in insulators, optical excitations typically observed are holons, doublons, and excitons. Excitons are particle-hole pairs bounded by Coulomb interaction and can be classified depending on the strength of such interaction~\cite{Fox2010,Wooten2013}.\ \emph{Wannier-Mott} excitons are realized when the particle-hole interaction is weak, giving rise to a large-radius quasiparticle. These excitons are typically found in semiconductors.\ \emph{Frenkel} excitons are tightly bound with a small particle-hole radius. Conjugated polymers and molecular crystals usually display these excitons.\ \emph{Mott-Hubbard} excitons arise from local Coulomb interactions such as Hubbard-$V$ terms, and they have been proposed to explain some optical properties in complex oxides.

Melting of insulating behavior with strong ultrashort light pulses has been experimentally explored. Ultrafast photodoping in the one-dimensional Mott insulator ET-$\mathrm{F_2\, TCNQ}$ has induced a metastable metallic behavior~\cite{Wall11}. It is also possible to control the generation of Mott-Hubbard excitons and their dynamics~\cite{Mitrano14}. Similarly, reversible control over the dielectric breakdown in fused silica has lead to a current increase of over 18 orders of magnitude~\cite{Schiffrin13,Schultze13}.

Photoinduced states in insulator complex-oxides, where orbitals play a role, have also been studied. An insulator-metal phase transition generated with light in $\mathrm{Pr_{0.7}Ca_{0.3}MnO_3}$ was reported~\cite{Polli07}. Using optical reflectivity measurements, the emergence of a metallic phase, melting of charge and orbital order, and coherent orbital oscillations were demonstrated. This photometal is not present in the phase diagram of such a compound. Analogously, a photoinduced insulator-ferromagnet phase transition was observed in strained $\mathrm{La_{2/3}Ca_{1/3}MnO_3}$~\cite{Zhang16}.

Theoretical effort has been devoted to understanding the breakdown of single-orbital Mott insulators~\cite{Oka03,Oka12,Khaled08,Fabian10,Hofmann12,Eck10,Eck13,Kanamori09,Maeshima10prb,Kanamori11,Kanamori12,Ishihara13}. The key finding in these efforts is the Mott insulator transitions to a quasistationary bad-metal state, which survives due to Landau-Zener quantum tunneling between many-body states. The breakdown depends not only on the field strength but also on the Hubbard-$U$ interaction. Light-induced phenomena in two-orbital models have been studied focusing mainly on quarter-filled Mott insulators using double-exchange models~\cite{Kanamori09,Maeshima10prb,Kanamori11,Kanamori12,Ishihara13}. A field-generated high-low spin insulator-ferromagnet transition was observed, as well as bound states, and orbital coherent oscillations. 

Here, using time-dependent density matrix renormalization group (DMRG)~\cite{dmrg1,*dmrg2,dmrg3,*dmrg4,White04,Daley04,Feiguin05,rk4}, we study the breakdown of a Mott insulator in the half-filled two-orbital Hubbard model. We show that this mechanism is strongly enhanced by the Hund interaction, melting the antiferromagnetism (AFM) and giving rise to an unconventional type of excitations that we dub ``Hund excitons''. Unlike those previously discussed, these are not driven by Coulomb interactions but by Hund's exchange. We also show that Hund interaction \emph{dynamically} activates large orbital fluctuations in the field-induced dynamics.

\section{Theoretical considerations\label{sec:theo}}
In this section we describe the model Hamiltonian and parameters used in our study, and the observables calculated to analyze the photodynamics in the two-orbital Mott insulator. We also discuss some basic properties of the half-filled Mott insulating state.

\subsection{Model Hamiltonian}
We explore a two-orbital Hubbard model Hamiltonian which includes kinetic energy, intra- and inter-orbital local Coulomb repulsion, Hund's rule exchange coupling, and pair-hopping process. Explicitly,
\begin{align}
H =-&\sum_{i\sigma\gamma} t_{\gamma} \Bigl( c^\dagger_{i \sigma \gamma}c_{i+1 \sigma \gamma} + \mathrm{H.c.} \Bigr) + U \sum_{i\gamma} n_{i\uparrow\gamma}n_{i\downarrow\gamma} \nonumber \\
-&\, 2J\sum_{i} \mathbf{S}_{i1} \cdot \mathbf{S}_{i2} + \left(U' - J/2\right)\sum_{i} n_{i1}n_{i2} 
\label{eq:H} \\
+&\, J\sum_{i} \left( c^\dagger_{i\uparrow 1}c^\dagger_{i\downarrow 1} c_{i\downarrow 2}c_{i\uparrow 2} + \mathrm{H.c.} \right) \nonumber,
\end{align}
where we consider a one-dimensional geometry of $N$ sites, or equivalently $2N$ orbitals, whose sites are labeled by $i$. The operator $c_{i \sigma \gamma}~(c^\dagger_{i \sigma \gamma})$ annihilates (creates) a particle at site $i$, with a spin projection $\sigma$ at orbital $\gamma = 1,\, 2$. The density $n_{i\sigma\gamma}$ and spin $\mathbf{S}_{i\gamma}$ operators are written in the standard form in terms of $c_{i \sigma \gamma}$ and $c^\dagger_{i \sigma \gamma}$. The hopping parameters $t_\gamma$ are orbital-dependent. $U$ and $J$ stand for Hubbard repulsion and Hund's ferromagnetic exchange, and $U' = U - 2J$ is used for symmetry reasons.

\subsection{Light pulse}
To account for the interaction with a light pulse, we employ the so-called Peierls substitution, where the \textit{classical} light field enters in the Hamiltonian through the kinetic energy operator. We therefore introduce a time $\tau$ dependence in the hopping integrals:
\begin{equation}
t_\gamma \longrightarrow t_\gamma(\tau) = t_\gamma \exp(i A(\tau) / N),
\end{equation}
where $N$ stands for the system size. The time-dependent phase is actually the vector potential of light whose form is assumed to be an oscillatory Gaussian pulse
\begin{equation}
A(\tau) = A_0 \exp \left[ -\frac{(\tau - \tau_P)^2}{2\sigma_P^2} \right] \cos \left[ \omega_P (\tau - \tau_P) \right].
\end{equation}
The frequency $\omega_P$, intensity $A_0$, width $\sigma_P$, and peak time $\tau_P$ characterize the nature of the electric field of the light pulse, for which we neglect its dynamics and assumed to be a classical field. In the calculations presented here, we have only considered few-cycle pulses given its experimental relevance.

\subsection{Parameters and procedure}
The procedure followed for the calculation of the time-dependent properties of the two-orbital Hubbard model is as follows. We first calculate the ground state of Hamiltonian~(\ref{eq:H}), without light field ($A(\tau) = 0$), using ground-state DMRG~\cite{dmrg1,*dmrg2,dmrg3,*dmrg4}. Once the energy has been minimized and the wavefunction $|\Psi_0\rangle$ is accurately represented, we switch on the vector potential and perform a real-time evolution of the ground state using the now time-dependent Hamiltonian~(\ref{eq:H}) and solving for the wavefunction through the equation $|\Psi(\tau + d\tau)\rangle = \exp( - i \int_{\tau}^{\tau + d\tau} d\tau' H( \tau') ) |\Psi(\tau')\rangle$, with initial condition $|\Psi(\tau = 0)\rangle = |\Psi_0\rangle$. The time evolution is performed using time-dependent DMRG with a fourth-order Runge-Kutta integrator to adapt the basis~\cite{Feiguin05,rk4,dmrg3,*dmrg4}, with a Krylov expansion of the evolution operator~\cite{rk4}. With $|\Psi(\tau)\rangle$ for different time slices, we can calculate expectation values and monitor their time evolution. The DMRG simulations were performed for systems from 8 up to 32 orbital sites with a discarded weight of less than $10^{-8}$. We note that the finite-size effect in this model has been shown to be quite small. Typically, for $N \gtrsim 8$, such effects are negligible for the purpose of our investigations.

The parameters of the Hamiltonian to be used are $t_\gamma = (-0.5, -0.5)$ and $U = 8$, with a filling factor of one particle per orbital, i.e.~half-filling, and we will explore results for the ratios $J/U = 0,\, 0.1,\, 0.25$. We have chosen a diagonal hopping matrix merely for simplicity, based on previous work~\cite{Rincon14prl,Rincon14}. Materials such as the iron-based superconductors have $J/U$ as large as 1/4~\cite{Dagotto12}. The bandwidth for each orbital is $W_\gamma = 4 t_\gamma$. We will refer to the case when $W_1 = W_2$ as \emph{isotropic}. The name is related to the presence of a $\mathrm{U}(1)_o$ symmetry associated to the orbital channel. For the isotropic case the full symmetry of the system is $\mathrm U(1)_c \times \mathrm{SU}(2)_s \times \mathrm{U}(1)_o$, where the subindices refer to charge conservation $(c)$, spin rotational symmetry $(s)$, and orbital rotational symmetry $(o)$. The parameters used for the light field are $A_0 = 2.4$, $\sigma_P = 0.19$, $\tau_P = 4\sigma_P$, finally, $\omega_P = \Delta$ the frequency of the pulse was set to match the charge gap $\Delta$ of the Mott insulator ground state.

\subsection{Observables}
In order to study the time response of the system we explore the following observables. The total double occupancy
\begin{equation}
D_{\rm tot}(\tau) = \frac{1}{N} \sum_i \langle \Psi(\tau) | n_{i\uparrow} n_{i\downarrow} | \Psi(\tau) \rangle.
\end{equation}
The local magnetic moment
\begin{equation}
\langle S^z(\tau)^2 \rangle = \frac{1}{N} \sum_i \langle \Psi(\tau) | (S^z_i)^2 | \Psi(\tau) \rangle.
\end{equation}
The orbital-dependent static spin structure factor
\begin{equation}
S_\gamma(q, \tau) = \frac{1}{N} \sum_{jk} e^{iq(j-k)} \langle \Psi(\tau) | S^+_{j\gamma}S^-_{k\gamma} | \Psi(\tau) \rangle.
\end{equation}
Inter-orbital local charge fluctuations are
\begin{equation}
C_{12}(\tau) = \frac{1}{N} \sum_i \langle \Psi(\tau) | \delta n_{i1} \delta n_{i2} | \Psi(\tau) \rangle,
\end{equation}
where $\delta n_{i\gamma} := n_{i\gamma} - \langle \Psi(\tau) | n_{i\gamma} | \Psi(\tau) \rangle$. The local holon $\left[ n^h_{i\gamma} := (1 - n_{i\uparrow\gamma})(1 - n_{i\downarrow\gamma}) \right]$ doublon ($n^d_{i\gamma} := n_{i\uparrow\gamma} n_{i\downarrow\gamma}$) number correlation function is defined as
\begin{equation}
C^{dh}_{\rm loc}(\tau) = \frac{1}{N} \sum_i \langle \Psi(\tau) | n^d_{i1} n^h_{i2} + n^h_{i1} n^d_{i2} | \Psi(\tau) \rangle.
\end{equation}
Similarly the orbital nearest-neighbor doublon-holon correlation function is
\begin{equation}
C^{dh}_\gamma(\tau) = \frac{1}{N-1} \sum_i \langle \Psi(\tau) | n^d_{i\gamma} n^h_{i+1\gamma} + n^h_{i\gamma} n^d_{i+1\gamma} | \Psi(\tau) \rangle.
\end{equation}

In the plots shown in this paper the ordinate axis actually represents the relative change of any of the above mentioned expectation values with respect to their ground-state value. That is, any expectation value $\mathcal E_{a}(\tau)$, which depends on time and has quantum numbers packed as a single index $a$, will be plotted as
\begin{equation}
\Delta \mathcal E_{a}(\tau) := \frac{\mathcal E_{a}(\tau) - \mathcal E_{a}(\tau = 0)}{\mathcal E_{a}(\tau = 0)}.
\end{equation}
For simplicity we will label the ordinate axis by $\mathcal E_{a}(\tau)$ instead of $\Delta \mathcal E_{a}(\tau)$. For instance, for the total double occupancy, we will plot
\begin{equation}
\Delta D_{\rm tot}(\tau) := \frac{D_{\rm tot}(\tau) - D_{\rm tot}(0)}{D_{\rm tot}(0)},
\end{equation}
but label the ordinate axis in the plots as just $D_{\rm tot}(\tau)$. An important exception will be the expectation value of the current density operator defined as
\begin{equation}
J(\tau) = \frac{i}{N-1} \sum_{j\sigma\gamma} t_{\gamma} \langle \Psi(\tau) | c^\dagger_{j \sigma \gamma}c_{j+1 \sigma \gamma} - \mathrm{H.c.} | \Psi(\tau) \rangle,
\end{equation}
for $\tau > 2 \tau_P$, which we will just simply plot as $J(\tau)$.

\subsection{On the nature of the ground state}
Let us first discuss the nature of the ground state of the two-orbital Hubbard model at half-filling. Regardless of the values of $U / W_\gamma$, the system has a finite single-particle gap, signaling an insulating state. In the absence of Hund's coupling, the ground state is weakly magnetic with no large AFM and large orbital fluctuations. For finite values of $J$, the ground state has a local magnetic moment leading to the establishment of quasi-long-range antiferromagnetic correlations with a large suppression of fluctuations in the orbital channel. The imbalance in the bandwidths of the orbitals will still favor antiferromagnetic order and, under the appropriate conditions, the orbital differentiation phenomena~\cite{Georges13}.

\section{Results for the photodynamics\label{sec:res}}
\begin{figure}
\centering
\includegraphics*[height=4.5cm]{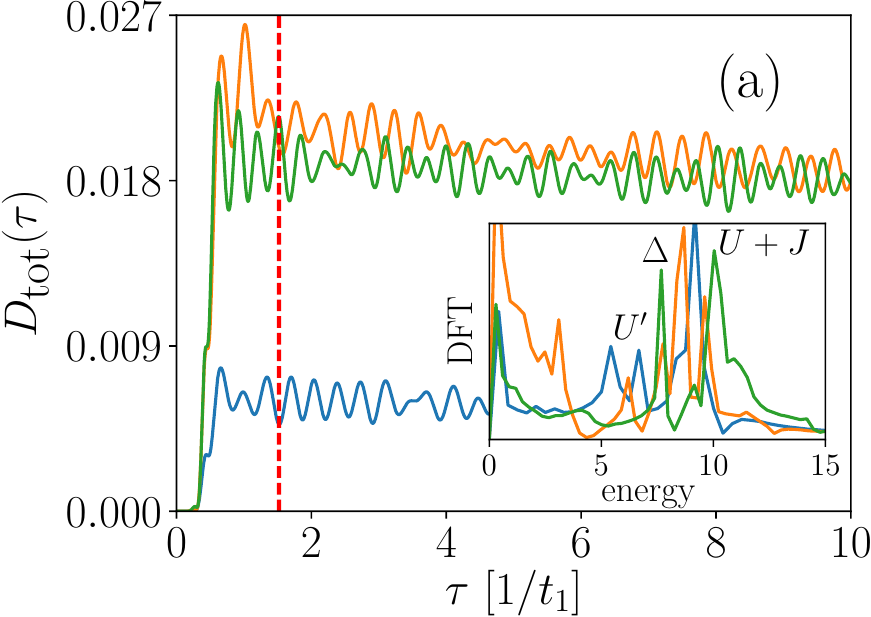}
\includegraphics*[height=4.6cm]{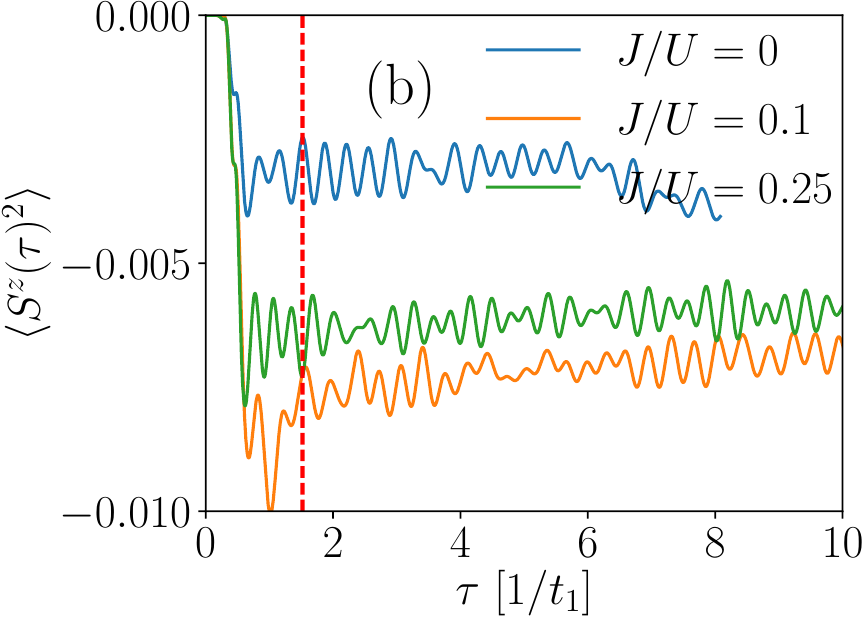}
\caption{Relative double occupancy (a) and local magnetic moment (b) for several values of $J/U$ and $U = 4W$. The vertical dashed line marks the light pulse span. (a) Breakdown of the Mott insulator is signaled by the creation of doublons. Inset: discrete Fourier transform (DFT) of $D_{\rm tot}(\tau)$, color coded with the main figure, showing the characteristic energy scales associated with the oscillations in the time response and the corresponding peak-splitting effect of $J$. (b) A decrease in the local magnetic moment indicates a high- to low-spin transition induced by the field.}
\label{fig:dd}
\end{figure}

\begin{figure}
\centering
\includegraphics*[height=4.5cm]{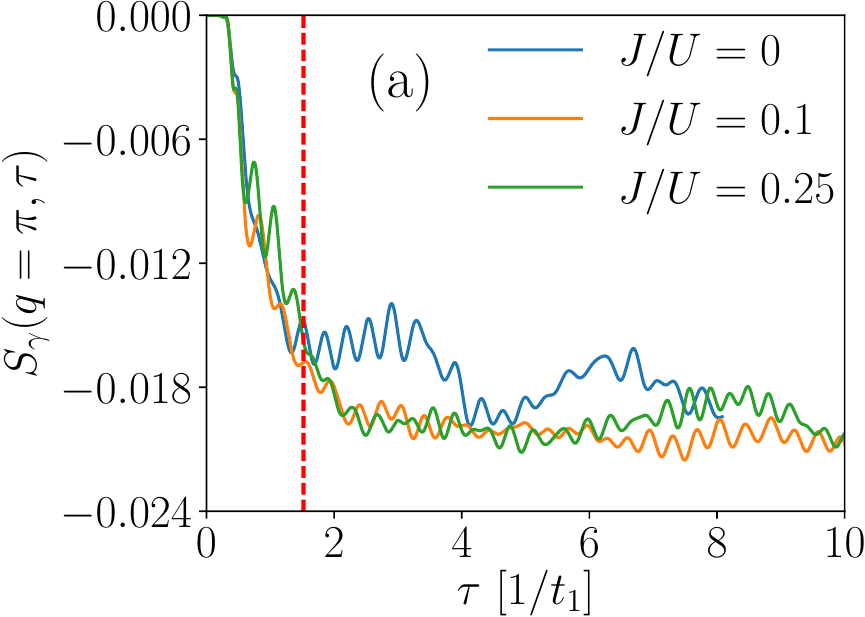}
\includegraphics*[height=4.3cm]{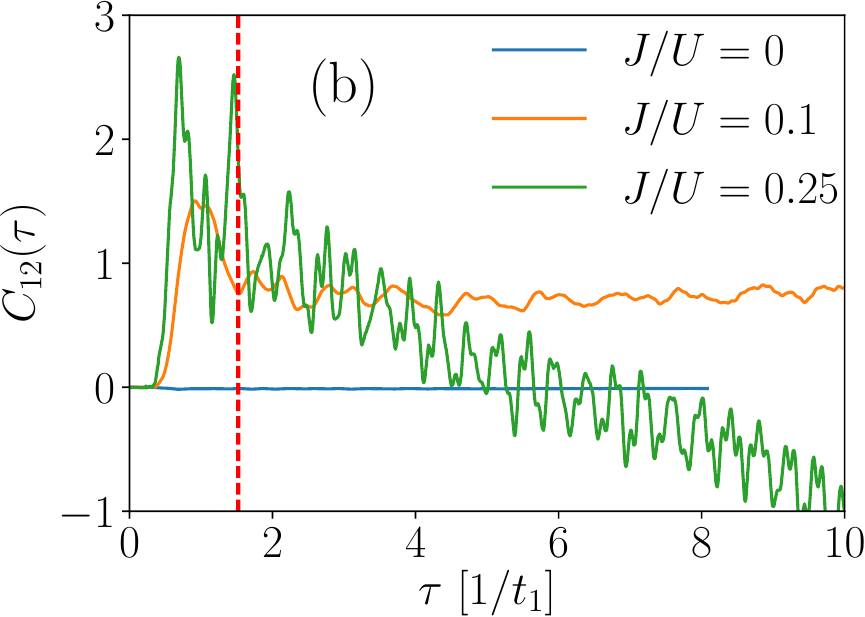}
\caption{Relative antiferromagnetic fluctuations at $q = \pi$ (a) and inter-orbital local charge fluctuations (b), for several values of $J/U$. The vertical dashed line marks the light pulse span. (a) The AFM order is partially melted within a timescale different than that of doublon production. (b) The dynamical coupling effect of Hund's exchange can be seen in the interorbital charge fluctuations.}
\label{fig:mm}
\end{figure}

\begin{figure}
\centering
\includegraphics*[height=4.5cm]{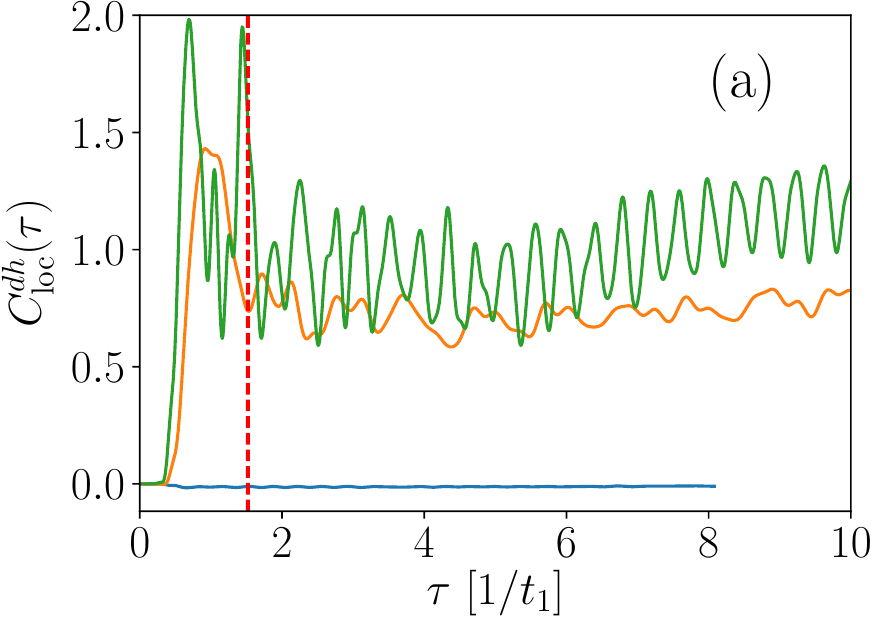}
\includegraphics*[height=4.5cm]{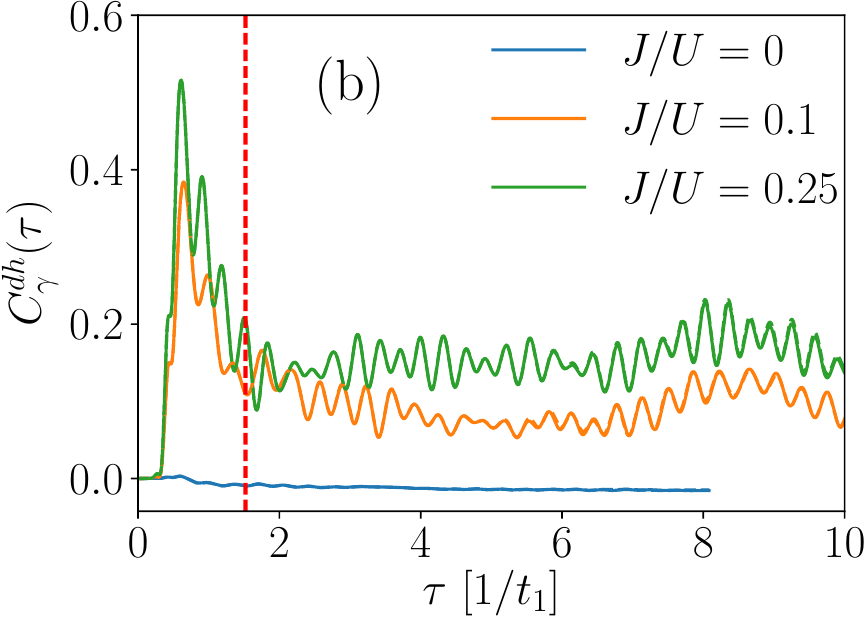}
\caption{Local (a) and neighboring (b) doublon-holon correlation function for several values of $J/U$. The vertical dashed line marks the light pulse span. Large fluctuations in the local doublon-holon number correlation (a), when compared to the neighboring one (b), suggest that doublon and holon will tend to bound on the same site and different orbitals rather than within neighboring sites in the same orbital.}
\label{fig:dh}
\end{figure}

Results for the photodynamics are shown in Figs.~\ref{fig:dd}-\ref{fig:cc}. We first consider $J=0$. The ground state in this case is paramagnetic, which implies that the doublon number is large (compared to the case $J \neq 0$). The large amount of doublons in the ground state has to do with small magnetic and large orbital fluctuations; therefore, the resulting light-induced doublon production is weak. As shown in Fig.~\ref{fig:dd}~(a), the effective photodoping in this case (measured as the amount of doublon or holons generated with respect to the ground state) is about $0.6\%$. The periods of the oscillations are $2\pi/U$ and $2\pi/\Delta$, where $\Delta$ corresponds to the single-particle gap.

In contrast, the $J \neq 0$ case leads to a larger production of doublon-holon pairs and hence to a stronger breakdown of the insulating state, see Fig.~\ref{fig:dd}~(a). (The breakdown of the insulator is to be understood here as the production of doublon-holon pairs, which constitute the carriers.) We observe that for fairly large values of $J/U$ the photodoping rate lies around $2\%$. We can then say that Hund's exchange enhances the melting of the Mott insulator. Notice that finite $J$ lifts the degeneracy of the atomic states giving rise to periods of oscillation $2\pi/U'$, $2\pi/\Delta$, and $2\pi/(U+J)$. We see that the effective Hubbard $U$ has been renormalized by $J$. As we will discuss below, the latter period is associated to the photogeneration of spin-singlet orbital-triplet doublons.

The production of carriers can be associated with a reduction in the local magnetic moment. The time evolution of $\langle S^z(\tau)^2 \rangle$, shown in Fig.~\ref{fig:dd}~(b), indicates a decrease in the magnetic moment. Such change agrees with both doublon production and melting of magnetic order signaling a high- to low-spin photoinduced phase transition. The reduction in $\langle S^z(\tau)^2 \rangle$ is only partial and so is the phase transition between spin states, where $\langle S^z(\tau)^2 \rangle$ has been reduced by up to $1\%$.

The doublon-holon pair production leads to a concomitant partial melting of the magnetic order. We plot in Fig.~\ref{fig:mm}~(a) the value of the spin structure factor at the ordering wave vector $q/\pi = 1$. We observe a cooperative phenomenon where the melting of the AFM order happens simultaneously as doublons are produced. The typical time scale at which $S_\gamma(q=\pi, \tau)$ decays is related to the single-particle charge gap $\Delta$, which in the atomic limit goes as $\Delta^{\rm at} \sim U+J$ such that the decrease in the AFM fluctuations occurs at larger time scales for finite $J$.

In order to examine the photodynamics in the orbital channel we show in Fig.~\ref{fig:mm}~(b) the change in the orbital fluctuations, $C_{12}(\tau)$, as a function of time $\tau$. We observe this change to be negligible for $J = 0$. The situation completely changes when $J/U \neq 0$, where large variations in $C_{12}$ are detected. We also notice that larger values of $J/U$ lead to larger fluctuations. The negative deviations for $J/U = 1/4$ show that orbital fluctuations at long times decrease compared to the ground-state value, still indicating a substantial change in orbital correlations. The positive-to-negative change in $C_{12}(\tau)$ occurs due to the large value of the ratio $J/U = 1/4 \lesssim 1/3$. Indeed, large values of $J$ will have the tendency to primarily modify orbital correlations. 

The light-induced activation of correlations in the orbital channel, as seen in the time dependence of $C_{12}(\tau)$, reaches an order-of-magnitude increase when compared to the ground-state case. At equilibrium, Hund's coupling tends to lock orbital correlations leading to the formation of robust magnetic moments and order; whereas in the nonequilibrium case, Hund's exchange \emph{dynamically} favors orbital excitations and melting of such magnetic moments. The activation of orbital fluctuations is due to the suppression of the magnetic order and corresponding generation of doublons, which can hop through the system via second-order- and pair-hopping processes. This makes the coupling of orbitals stronger than in the ground-state scenario.

The partial melting of AFM, the high-low spin photoinduced transition, and the large orbital fluctuations (with the corresponding oscillation frequency $U+J$ associated to spin-singlet orbital-triplet doublons) indicate that the doublons and holons arrange in an unexpected way in the photodynamics. We confirm this by studying the local interorbital $C^{dh}_{\rm loc}$ and neighboring intraorbital $C^{dh}_{\gamma}$ correlations between holon and doublon number operators, see Fig.~\ref{fig:dh}. We observe that the local doublon-holon number correlation displays robust changes, as large as $60\%$ up to $150\%$, compared to the ground state [Fig.~\ref{fig:dh}~(a)]. On the other hand, the neighboring intraorbital doublon-holon number correlation presents smaller changes, of around $10-20\%$, compared to $C^{dh}_{\rm loc}$ [Fig.~\ref{fig:dh}~(b)]. We can clearly see the effect of Hund's coupling in the photodynamics: without $J$ there is no appreciable change in both $C^{dh}_{\rm loc}$ and $C^{dh}_{\gamma}$, and large values of $J/U$ lead to a larger response in said observables. The large effect in $C^{dh}_{\rm loc}(\tau)$ is associated to pair- and second-order-hopping processes along with spin fluctuations, which do not necessarily enhance $C^{dh}_{\gamma}(\tau)$. At equilibrium, Hund's coupling locks orbital correlations leading to robust magnetic moments and order~\cite{Georges13,Rincon14prl}; in the nonequilibrium case, Hund's exchange \emph{dynamically} favors orbital excitations and melting of such moments.

\subsection{Exciton properties}
\begin{figure}
\centering
\includegraphics*[width=.85\columnwidth]{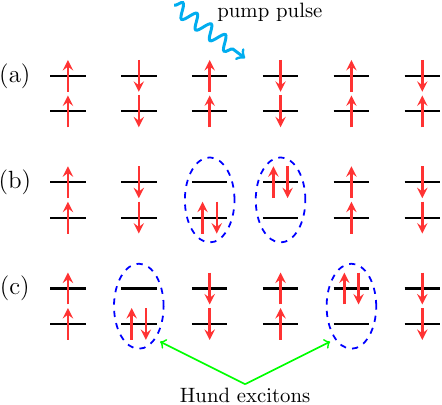}
\caption{Dynamical formation of excitons in a one-dimensional two-orbital Mott insulator. (a) The starting point is the Mott insulator ground state interacting with the incident light. (b) After photoirradiation, doublon-holon pairs are created by absorption of photons, which bound to form Hund excitons (dashed ellipses). (c) Such excitons will move in the lattice with a characteristic bandwidth~(\ref{eq:bw}) and size~(\ref{eq:es}). While in principle there are other possible states that can be generated from panel (a) after the absorption of a photon, our results indicate that panel (b) is the most likely.}
\label{fig:excGen}
\end{figure}

We interpret the above mentioned results as the \emph{photogeneration of neutral low-spin objects with orbital character that are bound locally}, suggesting they are unconventional excitons. More specifically, the photodynamics generates spin-singlet orbital-triplet doublon-holon bound pairs induced by $J$. These excitations move in an AFM background with an effective bandwidth
\begin{equation}
W_{\rm excitons} \sim
\dfrac{t_1 t_2}{U - 3J}.
\label{eq:bw}
\end{equation}
Naively, it is expected that the excitons, which are naturally Bosons, will propagate coherently giving rise to a condensate in the low-density limit. These ``Hund excitons'' are created due to the presence of Hund's exchange; therefore their existence is \emph{not} directly related to Coulomb interactions. These findings suggest that the type of excitons observed here correspond to a kind of optical excitation that is set apart from more familiar quasiparticles, such as Frenkel, Mott-Hubbard, and Wannier-Mott excitons, which are typically associated to large or small Coulomb interactions. A cartoon showing the photogeneration of Hund excitons and their emergent dynamics is shown in Fig.~\ref{fig:excGen}.

\begin{figure}
\centering
\includegraphics{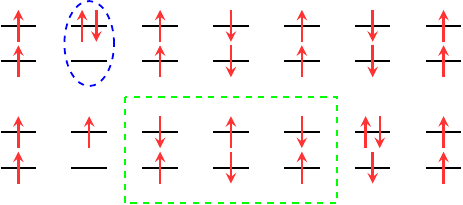}
\caption{Cartoon illustrating the stability of Hund excitons in a one-dimensional two-orbital model at half-filling in the Ising limit (ignoring spin fluctuations). The AFM background induces the linear confining potential~(\ref{eq:cp}). Starting with a single exciton (top) and subsequently moving the doublon (or analogously the holon) will distort both the local magnetic moment and the AFM background by destroying local triplet states and reducing super-exchange (bottom, dashed box). The magnetic energy loss, which is proportional to the length of the box or number of hops, will generate the trapping potential~(\ref{eq:cp}) for the doublon (or analogously, for the holon).}
\label{fig:confPot}
\end{figure}

The stability of the excitons can be understood by considering the two-orbital AFM insulating state in the Ising limit (ignoring spin fluctuations) and assuming the creation of a single doublon-holon pair locally, as shown in Fig~\ref{fig:confPot}. For illustration, one can allow the doublon to move through hopping processes, creating a string of misaligned spins in the AFM background, separating it from the holon (see Fig.~\ref{fig:confPot}). This will induce a linear confining potential $V_{\rm excitons}(x)$, at position $x$, that will favor the motion of both holon and doublon as a single ``heavier'' object, with the bandwidth~(\ref{eq:bw}). The confining potential reads
\begin{equation}
V_{\rm excitons}(x = ja) \sim J | x |,
\label{eq:cp}
\end{equation}
where $J$ corresponds to Hund's coupling, $j$ labels a site, and $a$ is the lattice spacing. In order to arrive to this result, we have only considered the Hund exchange in the $z$ direction. Notice that the existence of the confining potential is directly linked to a finite value of Hund's rule coupling. If $J = 0$, there is no confining potential, allowing the doublons and holons to move independently.

We can also estimate the size of the exciton by writing down an effective model for the motion of a doublon (or a holon) in the presence of the exciton potential~(\ref{eq:cp}). The effective Hamiltonian is
$
H_{\rm eff} = -\sum_m t_0 |m\rangle\langle m+1| + t_0 |m+1\rangle\langle m| - V_{\rm excitons}(m) |m\rangle\langle m|,
$ 
where $t_0 = \max(t_1, t_2)$ and $|m\rangle$ correspond to a Wannier orbital at lattice site $m$. Writing down the wave function of the doublon as $|\mathrm{doublon}\rangle = \sum_m c_m |m\rangle$ we obtain the following time-independent Schr\"odinger equation:
\begin{equation}
Ec_m = -t_0 ( c_{m-1} + c_{m+1}) + V_{\rm excitons}(m) c_{m},
\end{equation}
where $E$ stands for the energy. Taking the continuum limit of this equation one obtains
\begin{equation}
-t_0 \frac{d^2 c_x}{dx^2} + V_{\rm excitons}(x) c_x = E c_x.
\end{equation}
And by re-expressing it in terms of dimensionless variables, one obtains the size of the excitons as the characteristic length scale of the problem:
\begin{equation}
\ell_{\rm excitons} \sim a \left( \frac{t_0}{J} \right)^{1/3}.
\label{eq:es}
\end{equation}

For the parameters considered in this work we find that $\ell_{\rm excitons}/a \approx 2^{-2/3} \approx 0.63$ for $J/U = 0.25$, and $\ell_{\rm excitons}/a \approx 5^{1/3}/2 \approx 0.86$ for $J/U = 0.10$, showing that the bound state is essentially local with almost no spatial fluctuations. A more detailed analysis beyond the semiclassical approximation made here will be done in an upcoming work.

The physics of photogeneration of Hund excitons somehow resembles the situation in a slightly doped two-dimensional AFM. There, the holon is trapped in a linear potential created by the AFM background~\cite{Dagotto94}. Similar confinement features between holes owed to an interplay of superexchange and delocalization were explored in a two-band Hubbard model~\cite{Khaled09}. In the case considered here, holon and doublon couple locally in different orbitals in order to minimize $V_{\rm excitons}(x)$ so that the AFM established by Hund's coupling and superexchange is minimally disturbed. The effect of confinement of spinons, via a linear potential induced by interchain coupling, has been observed experimentally in the spin systems $\mathrm{XCo_2V_2O_8}$ ($\mathrm{X = Ba, Sr}$)~\cite{Grenier15,Bera17,Faure17}. The confinement occurs in the spin channel between different chains; in our case it appears in the charge channel between different orbitals.

The experimental detection of Hund excitons can be accomplished, for example, in a pump-probe (time-resolved reflectivity) setup, by registering the evolution of the optical conductivity as a function of both energy and delay time. A resulting midgap or midband state, depending on $U/W$ and $J/U$, should be detected between an optical band of unbounded doublon-holon pairs and a precursor of the Drude peak. (There might be multiparticle absorption bands at higher energies that are not relevant to the analysis here.) Such midgap/midband peak will correspond to the excitons analyzed here. These excitons could be observed in oxide perovskites, such as manganites and nickelates, where two orbitals are needed for their description.

\begin{figure}
\centering
\includegraphics*[height=4.6cm]{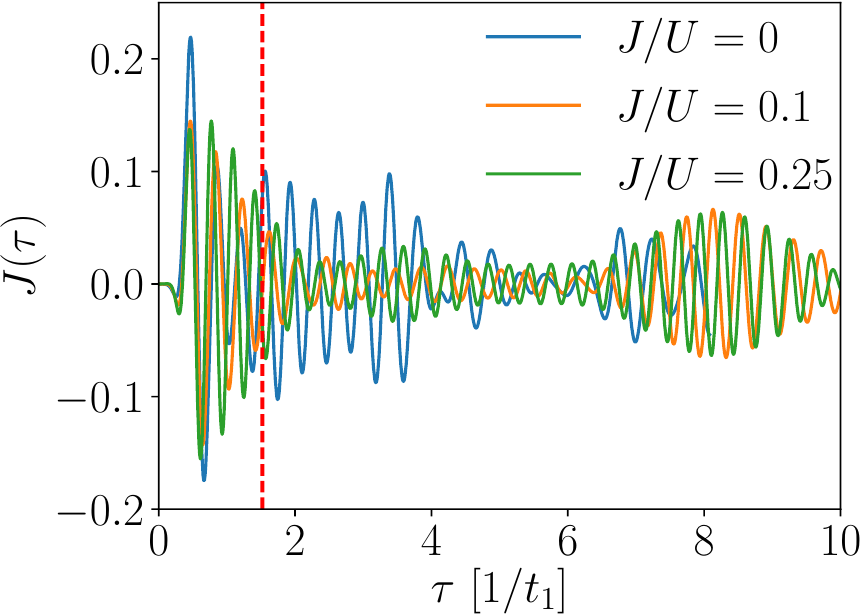}
\caption{Electric current density averaged over the system size, for several values of $J/U$. The vertical dashed line marks the light pulse span. A long-lived metallic state with a small, albeit finite, oscillating current is found. The photocarriers correspond to a plasma of doublon-holon pairs coexisting with Hund excitons.}
\label{fig:cc}
\end{figure}

The photocurrent resulting from the breakdown of the Mott insulator is displayed in Fig.~\ref{fig:cc}. We observe a transient dynamics, from a high-current state right after the pulse is applied to a low-current steady state. We note that $J(\tau)$ displays a similar trend to the intraorbital correlation $C^{dh}_{\gamma}(\tau)$, indicating a synergy between intraorbital doublon-holon pairs and photocarrier dynamics. The oscillations in $J(\tau)$ are associated to frequencies related to $U'$ and $\Delta$. We notice that the revival of the current at later times is due to the system's finite size (see below). The fact that the generation of doublons (Fig.~\ref{fig:dd}), and therefore Hund excitons, flattens out at late times and the fact that we have a stationary current implies that the excitons coexist with a plasma of mobile carriers (dissociated doublon-holon pairs). In other words, Hund excitons are created up to a saturation point and then the metallic state follows due to their partial dissociation. This in contrast with previous results in single-orbital Hubbard models~\cite{Eck10}, where a constant creation of doublons has no impact on the current. 

As shown by the results in Figs.~\ref{fig:dd}-\ref{fig:cc}, the photoinduced state corresponds to an out-of-equilibrium strongly correlated metal with partially polarized spins and orbital-dependent correlations, with the concurrency of Hund excitons, and doublon-holon plasma as photocarriers. A possible scenario that could be realized at long times is that the current completely decays to zero, leading to the possibility of a photoinduced phase transition to a Hund-excitonic insulator. A separate future study on the role of fluence and system size will clarify whether such feature survives in the long-time photodynamics.

Finally, let us comment on the robustness of the Hund exciton phenomenology. It has been shown~\cite{Oka03,Oka12,Khaled08,Fabian10,Hofmann12,Eck10,Eck13,Kanamori09,Maeshima10prb,Kanamori11,Kanamori12,Ishihara13} that the main aspects of the two-dimensional phase diagram of the two-orbital model can be mimicked in one dimension. Additionally, previous calculations~\cite{Kanamori09,Maeshima10prb,Kanamori11,Kanamori12,Ishihara13} for the one-orbital case have shown that similar quantum dynamics can be found both in one and two dimensions. The two-orbital two-dimensional photo-dynamics of the Mott insulator was recently explored~\cite{Strand17}. The type of excitations observed is similar in nature to the Hund excitons reported in this work. The fact that the breakdown of the Mott insulator is dominated by the interactions, which are local on-site, i.e.~at the individual atomic level, lead us to believe that the Hund exciton phenomenon may be present in two and three dimensions.

\subsection{Finite-size effects}
\begin{figure}
\centering
\includegraphics*[height=4.6cm]{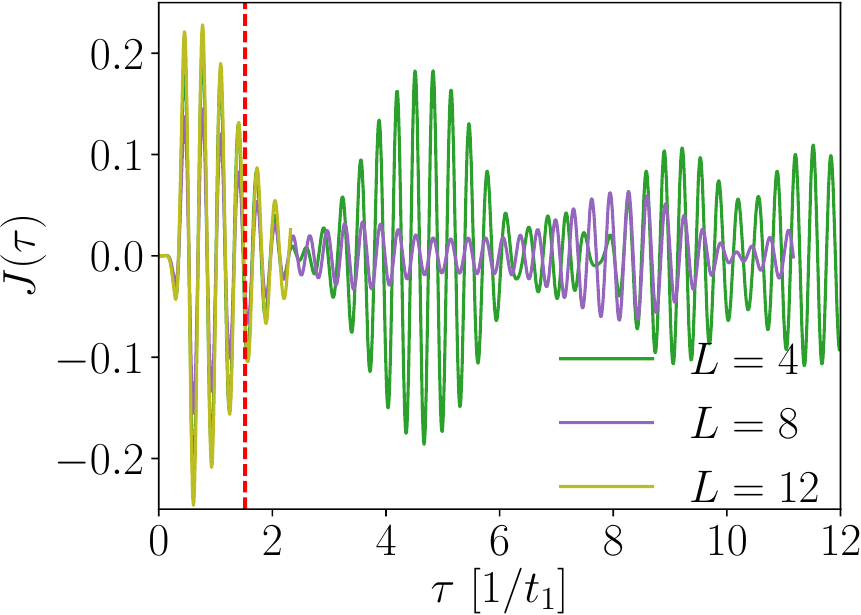}
\caption{Expectation value of the electric current density averaged over the system size, for several system sizes $L$ and for $J/U = 0.25$. The results for other values of $J/U$ are comparable. The vertical dashed line marks the light pulse span. The finite-size effects are virtually negligible at short times, which is when the photoexcitation of Hund excitons occurs. At intermediate and late times, size effects are stronger.}
\label{fig:finsize}
\end{figure}

In this section we briefly discuss the effect of calculating the time response of the two-orbital Mott insulator on a finite system. Such finite-size effects are more pronounced in the electric current. Figure~\ref{fig:finsize} shows the expectation value of the current density operator as a function of time $\tau$, for several system sizes $L = 4, 8, 12$ sites and $J/U = 0.25$. (Very similar results are found for other values of $J/U$.) The expectation value is averaged over the system size, see Sec.~\ref{sec:theo} for details, and time is scaled to the hopping parameter $t_1$.

It is possible to distinguish three different trends in this plot. At short times, the expectation value of the current density operator displays a fairly size-independent behavior. The values of $J(\tau)$ oscillate with the same period for all values of $L$. The values of those frequencies are related to the interorbital repulsion $U'$ and the charge gap $\Delta$, as was discussed in Sec.~\ref{sec:res}. For intermediate $\tau$, the values of the expectation value of the current operator still oscillate in phase; however, its strength changes greatly with the size of the system. The revival effect seen for $L = 4$ can therefore be attributed to finite-size effects. For late times, $J(\tau)$ oscillates out of phase and with a different strength for different values of $L$.

The light-assisted generation of Hund excitons and the accompanying phenomena discussed in this work, which is the main focus here, occur at early times. We therefore consider the influence of the above-mentioned finite-size effects as almost negligible, when describing the breakdown of the two-orbital Mott insulator.

\section{Conclusions\label{sec:con}}
We have studied the photoinduced melting of a two-orbital Mott insulator, resulting in a long-lived nonequilibrium metallic state with unexpected properties. We have found that the breakdown of the Mott insulator and corresponding onset of the photometallic state is enhanced by Hund's coupling. Indeed, Hund's exchange tends to suppress doubly-occupied configurations in the ground state, as opposed to the nonequilibrium situation where it greatly enhances doublon production. The breakdown of the insulating ground state is realized through an insulator-metal photoinduced phase transition with a concomitant melting of the magnetic moments, implying a high-low spin light-induced phase transition as well. Additionally, we have shown evidence that Hund's coupling \emph{dynamically} activates large orbital fluctuations leading to a nontrivial dynamics of doublon-holon pairs and to the loss of antiferromagnetic quasi-long-range order.

More importantly, we observed the dynamical formation of ``Hund excitons": emergent neutral quasiparticles with orbital and spin character stabilized by Hund's exchange, which can establish a condensate. In high contrast to the more familiar types of excitonic optical excitations, Hund excitons do not originated from direct Coulomb repulsion but from exchange interactions; in this case Hund's coupling. Moreover, we have studied semiclassically the properties of Hund excitons such as bandwidth, confining potential, and size as a function of Hund's exchange. We have also observed that the long-lived photometallic state results as a coexistence of Hund excitons and doublon-holon plasma.

It would be interesting to further explore the properties of Hund excitons. For example, the effect of exciton-exciton interactions on the photodynamics and the study of the stages of the breakdown of the Mott insulator. Indeed, a study on intraband relaxation dynamics in a two-band Mott insulator showed the generation of several excitations~\cite{Strand17}, among them Hund excitons. We note, however, that the coupling to radiation in that study differs from the one implemented here. These issues deserve further study and are currently under investigation.

\begin{acknowledgments}
J.R. acknowledges fruitful conversations with T. Oka, G. Baskaran, and Y. Wan. J.R. is supported by the Simons Foundation via the Many Electron Collaboration. A.E.F.~acknowledges the U.S.~Department of Energy, Office of Basic Energy Sciences, for support under grant DE-SC0014407. E.D.~was supported by the U.S. Department of Energy, Office of Basic Energy Sciences, Materials Sciences and Engineering Division. Numerical simulations were performed at the Center for Nanophase Materials Sciences, which is a DOE Office of Science User Facility. This research was supported in part by Perimeter Institute for Theoretical Physics. Research at Perimeter Institute is supported by the Government of Canada through the Department of Innovation, Science and Economic Development Canada and by the Province of Ontario through the Ministry of Research, Innovation and Science. 
\end{acknowledgments}

\bibliographystyle{apsrev4-1}
\bibliography{dielectric}

\begin{thebibliography}{45}%
\makeatletter
\providecommand \@ifxundefined [1]{%
 \@ifx{#1\undefined}
}%
\providecommand \@ifnum [1]{%
 \ifnum #1\expandafter \@firstoftwo
 \else \expandafter \@secondoftwo
 \fi
}%
\providecommand \@ifx [1]{%
 \ifx #1\expandafter \@firstoftwo
 \else \expandafter \@secondoftwo
 \fi
}%
\providecommand \natexlab [1]{#1}%
\providecommand \enquote  [1]{``#1''}%
\providecommand \bibnamefont  [1]{#1}%
\providecommand \bibfnamefont [1]{#1}%
\providecommand \citenamefont [1]{#1}%
\providecommand \href@noop [0]{\@secondoftwo}%
\providecommand \href [0]{\begingroup \@sanitize@url \@href}%
\providecommand \@href[1]{\@@startlink{#1}\@@href}%
\providecommand \@@href[1]{\endgroup#1\@@endlink}%
\providecommand \@sanitize@url [0]{\catcode `\\12\catcode `\$12\catcode
  `\&12\catcode `\#12\catcode `\^12\catcode `\_12\catcode `\%12\relax}%
\providecommand \@@startlink[1]{}%
\providecommand \@@endlink[0]{}%
\providecommand \url  [0]{\begingroup\@sanitize@url \@url }%
\providecommand \@url [1]{\endgroup\@href {#1}{\urlprefix }}%
\providecommand \urlprefix  [0]{URL }%
\providecommand \Eprint [0]{\href }%
\providecommand \doibase [0]{http://dx.doi.org/}%
\providecommand \selectlanguage [0]{\@gobble}%
\providecommand \bibinfo  [0]{\@secondoftwo}%
\providecommand \bibfield  [0]{\@secondoftwo}%
\providecommand \translation [1]{[#1]}%
\providecommand \BibitemOpen [0]{}%
\providecommand \bibitemStop [0]{}%
\providecommand \bibitemNoStop [0]{.\EOS\space}%
\providecommand \EOS [0]{\spacefactor3000\relax}%
\providecommand \BibitemShut  [1]{\csname bibitem#1\endcsname}%
\let\auto@bib@innerbib\@empty
\bibitem [{\citenamefont {Orenstein}(2012)}]{Orenstein12}%
  \BibitemOpen
  \bibfield  {author} {\bibinfo {author} {\bibfnamefont {J.}~\bibnamefont
  {Orenstein}},\ }\href {\doibase 10.1063/PT.3.1717} {\bibfield  {journal}
  {\bibinfo  {journal} {Physics Today}\ }\textbf {\bibinfo {volume} {65}},\
  \bibinfo {pages} {44} (\bibinfo {year} {2012})}\BibitemShut {NoStop}%
\bibitem [{\citenamefont {Kampfrath}\ \emph {et~al.}(2013)\citenamefont
  {Kampfrath}, \citenamefont {Tanaka},\ and\ \citenamefont
  {Nelson}}]{Kampfrath13}%
  \BibitemOpen
  \bibfield  {author} {\bibinfo {author} {\bibfnamefont {T.}~\bibnamefont
  {Kampfrath}}, \bibinfo {author} {\bibfnamefont {K.}~\bibnamefont {Tanaka}}, \
  and\ \bibinfo {author} {\bibfnamefont {K.~A.}\ \bibnamefont {Nelson}},\
  }\href {http://dx.doi.org/10.1038/nphoton.2013.184} {\bibfield  {journal}
  {\bibinfo  {journal} {Nat. Photon.}\ }\textbf {\bibinfo {volume} {7}},\
  \bibinfo {pages} {680} (\bibinfo {year} {2013})}\BibitemShut {NoStop}%
\bibitem [{\citenamefont {Fausti}\ \emph {et~al.}(2011)\citenamefont {Fausti},
  \citenamefont {Tobey}, \citenamefont {Dean}, \citenamefont {Kaiser},
  \citenamefont {Dienst}, \citenamefont {Hoffmann}, \citenamefont {Pyon},
  \citenamefont {Takayama}, \citenamefont {Takagi},\ and\ \citenamefont
  {Cavalleri}}]{Fausti11}%
  \BibitemOpen
  \bibfield  {author} {\bibinfo {author} {\bibfnamefont {D.}~\bibnamefont
  {Fausti}}, \bibinfo {author} {\bibfnamefont {R.~I.}\ \bibnamefont {Tobey}},
  \bibinfo {author} {\bibfnamefont {N.}~\bibnamefont {Dean}}, \bibinfo {author}
  {\bibfnamefont {S.}~\bibnamefont {Kaiser}}, \bibinfo {author} {\bibfnamefont
  {A.}~\bibnamefont {Dienst}}, \bibinfo {author} {\bibfnamefont {M.~C.}\
  \bibnamefont {Hoffmann}}, \bibinfo {author} {\bibfnamefont {S.}~\bibnamefont
  {Pyon}}, \bibinfo {author} {\bibfnamefont {T.}~\bibnamefont {Takayama}},
  \bibinfo {author} {\bibfnamefont {H.}~\bibnamefont {Takagi}}, \ and\ \bibinfo
  {author} {\bibfnamefont {A.}~\bibnamefont {Cavalleri}},\ }\href {\doibase
  10.1126/science.1197294} {\bibfield  {journal} {\bibinfo  {journal}
  {Science}\ }\textbf {\bibinfo {volume} {331}},\ \bibinfo {pages} {189}
  (\bibinfo {year} {2011})}\BibitemShut {NoStop}%
\bibitem [{\citenamefont {Miyano}\ \emph {et~al.}(1997)\citenamefont {Miyano},
  \citenamefont {Tanaka}, \citenamefont {Tomioka},\ and\ \citenamefont
  {Tokura}}]{Miyano97}%
  \BibitemOpen
  \bibfield  {author} {\bibinfo {author} {\bibfnamefont {K.}~\bibnamefont
  {Miyano}}, \bibinfo {author} {\bibfnamefont {T.}~\bibnamefont {Tanaka}},
  \bibinfo {author} {\bibfnamefont {Y.}~\bibnamefont {Tomioka}}, \ and\
  \bibinfo {author} {\bibfnamefont {Y.}~\bibnamefont {Tokura}},\ }\href
  {\doibase 10.1103/PhysRevLett.78.4257} {\bibfield  {journal} {\bibinfo
  {journal} {Phys. Rev. Lett.}\ }\textbf {\bibinfo {volume} {78}},\ \bibinfo
  {pages} {4257} (\bibinfo {year} {1997})}\BibitemShut {NoStop}%
\bibitem [{\citenamefont {Wall}\ \emph {et~al.}(2011)\citenamefont {Wall},
  \citenamefont {Brida}, \citenamefont {Clark}, \citenamefont {Ehrke},
  \citenamefont {Jaksch}, \citenamefont {Ardavan}, \citenamefont {Bonora},
  \citenamefont {Uemura}, \citenamefont {Takahashi}, \citenamefont {Hasegawa},
  \citenamefont {Okamoto}, \citenamefont {Cerullo},\ and\ \citenamefont
  {Cavalleri}}]{Wall11}%
  \BibitemOpen
  \bibfield  {author} {\bibinfo {author} {\bibfnamefont {S.}~\bibnamefont
  {Wall}}, \bibinfo {author} {\bibfnamefont {D.}~\bibnamefont {Brida}},
  \bibinfo {author} {\bibfnamefont {S.~R.}\ \bibnamefont {Clark}}, \bibinfo
  {author} {\bibfnamefont {H.~P.}\ \bibnamefont {Ehrke}}, \bibinfo {author}
  {\bibfnamefont {D.}~\bibnamefont {Jaksch}}, \bibinfo {author} {\bibfnamefont
  {A.}~\bibnamefont {Ardavan}}, \bibinfo {author} {\bibfnamefont
  {S.}~\bibnamefont {Bonora}}, \bibinfo {author} {\bibfnamefont
  {H.}~\bibnamefont {Uemura}}, \bibinfo {author} {\bibfnamefont
  {Y.}~\bibnamefont {Takahashi}}, \bibinfo {author} {\bibfnamefont
  {T.}~\bibnamefont {Hasegawa}}, \bibinfo {author} {\bibfnamefont
  {H.}~\bibnamefont {Okamoto}}, \bibinfo {author} {\bibfnamefont
  {G.}~\bibnamefont {Cerullo}}, \ and\ \bibinfo {author} {\bibfnamefont
  {A.}~\bibnamefont {Cavalleri}},\ }\href {http://dx.doi.org/10.1038/nphys1831}
  {\bibfield  {journal} {\bibinfo  {journal} {Nat. Phys.}\ }\textbf {\bibinfo
  {volume} {7}},\ \bibinfo {pages} {114} (\bibinfo {year} {2011})}\BibitemShut
  {NoStop}%
\bibitem [{\citenamefont {Schiffrin}\ \emph {et~al.}(2013)\citenamefont
  {Schiffrin}, \citenamefont {Paasch-Colberg}, \citenamefont {Karpowicz},
  \citenamefont {Apalkov}, \citenamefont {Gerster}, \citenamefont {Muhlbrandt},
  \citenamefont {Korbman}, \citenamefont {Reichert}, \citenamefont {Schultze},
  \citenamefont {Holzner}, \citenamefont {Barth}, \citenamefont {Kienberger},
  \citenamefont {Ernstorfer}, \citenamefont {Yakovlev}, \citenamefont
  {Stockman},\ and\ \citenamefont {Krausz}}]{Schiffrin13}%
  \BibitemOpen
  \bibfield  {author} {\bibinfo {author} {\bibfnamefont {A.}~\bibnamefont
  {Schiffrin}}, \bibinfo {author} {\bibfnamefont {T.}~\bibnamefont
  {Paasch-Colberg}}, \bibinfo {author} {\bibfnamefont {N.}~\bibnamefont
  {Karpowicz}}, \bibinfo {author} {\bibfnamefont {V.}~\bibnamefont {Apalkov}},
  \bibinfo {author} {\bibfnamefont {D.}~\bibnamefont {Gerster}}, \bibinfo
  {author} {\bibfnamefont {S.}~\bibnamefont {Muhlbrandt}}, \bibinfo {author}
  {\bibfnamefont {M.}~\bibnamefont {Korbman}}, \bibinfo {author} {\bibfnamefont
  {J.}~\bibnamefont {Reichert}}, \bibinfo {author} {\bibfnamefont
  {M.}~\bibnamefont {Schultze}}, \bibinfo {author} {\bibfnamefont
  {S.}~\bibnamefont {Holzner}}, \bibinfo {author} {\bibfnamefont {J.~V.}\
  \bibnamefont {Barth}}, \bibinfo {author} {\bibfnamefont {R.}~\bibnamefont
  {Kienberger}}, \bibinfo {author} {\bibfnamefont {R.}~\bibnamefont
  {Ernstorfer}}, \bibinfo {author} {\bibfnamefont {V.~S.}\ \bibnamefont
  {Yakovlev}}, \bibinfo {author} {\bibfnamefont {M.~I.}\ \bibnamefont
  {Stockman}}, \ and\ \bibinfo {author} {\bibfnamefont {F.}~\bibnamefont
  {Krausz}},\ }\href {http://dx.doi.org/10.1038/nature11567} {\bibfield
  {journal} {\bibinfo  {journal} {Nature}\ }\textbf {\bibinfo {volume} {493}},\
  \bibinfo {pages} {70} (\bibinfo {year} {2013})}\BibitemShut {NoStop}%
\bibitem [{\citenamefont {Schultze}\ \emph {et~al.}(2013)\citenamefont
  {Schultze}, \citenamefont {Bothschafter}, \citenamefont {Sommer},
  \citenamefont {Holzner}, \citenamefont {Schweinberger}, \citenamefont
  {Fiess}, \citenamefont {Hofstetter}, \citenamefont {Kienberger},
  \citenamefont {Apalkov}, \citenamefont {Yakovlev}, \citenamefont {Stockman},\
  and\ \citenamefont {Krausz}}]{Schultze13}%
  \BibitemOpen
  \bibfield  {author} {\bibinfo {author} {\bibfnamefont {M.}~\bibnamefont
  {Schultze}}, \bibinfo {author} {\bibfnamefont {E.~M.}\ \bibnamefont
  {Bothschafter}}, \bibinfo {author} {\bibfnamefont {A.}~\bibnamefont
  {Sommer}}, \bibinfo {author} {\bibfnamefont {S.}~\bibnamefont {Holzner}},
  \bibinfo {author} {\bibfnamefont {W.}~\bibnamefont {Schweinberger}}, \bibinfo
  {author} {\bibfnamefont {M.}~\bibnamefont {Fiess}}, \bibinfo {author}
  {\bibfnamefont {M.}~\bibnamefont {Hofstetter}}, \bibinfo {author}
  {\bibfnamefont {R.}~\bibnamefont {Kienberger}}, \bibinfo {author}
  {\bibfnamefont {V.}~\bibnamefont {Apalkov}}, \bibinfo {author} {\bibfnamefont
  {V.~S.}\ \bibnamefont {Yakovlev}}, \bibinfo {author} {\bibfnamefont {M.~I.}\
  \bibnamefont {Stockman}}, \ and\ \bibinfo {author} {\bibfnamefont
  {F.}~\bibnamefont {Krausz}},\ }\href {http://dx.doi.org/10.1038/nature11720}
  {\bibfield  {journal} {\bibinfo  {journal} {Nature}\ }\textbf {\bibinfo
  {volume} {493}},\ \bibinfo {pages} {75} (\bibinfo {year} {2013})}\BibitemShut
  {NoStop}%
\bibitem [{\citenamefont {Okamoto}\ \emph {et~al.}(2007)\citenamefont
  {Okamoto}, \citenamefont {Matsuzaki}, \citenamefont {Wakabayashi},
  \citenamefont {Takahashi},\ and\ \citenamefont {Hasegawa}}]{Okamoto07}%
  \BibitemOpen
  \bibfield  {author} {\bibinfo {author} {\bibfnamefont {H.}~\bibnamefont
  {Okamoto}}, \bibinfo {author} {\bibfnamefont {H.}~\bibnamefont {Matsuzaki}},
  \bibinfo {author} {\bibfnamefont {T.}~\bibnamefont {Wakabayashi}}, \bibinfo
  {author} {\bibfnamefont {Y.}~\bibnamefont {Takahashi}}, \ and\ \bibinfo
  {author} {\bibfnamefont {T.}~\bibnamefont {Hasegawa}},\ }\href {\doibase
  10.1103/PhysRevLett.98.037401} {\bibfield  {journal} {\bibinfo  {journal}
  {Phys. Rev. Lett.}\ }\textbf {\bibinfo {volume} {98}},\ \bibinfo {pages}
  {037401} (\bibinfo {year} {2007})}\BibitemShut {NoStop}%
\bibitem [{\citenamefont {Polli}\ \emph {et~al.}(2007)\citenamefont {Polli},
  \citenamefont {Rini}, \citenamefont {Wall}, \citenamefont {Schoenlein},
  \citenamefont {Tomioka}, \citenamefont {Tokura}, \citenamefont {Cerullo},\
  and\ \citenamefont {Cavalleri}}]{Polli07}%
  \BibitemOpen
  \bibfield  {author} {\bibinfo {author} {\bibfnamefont {D.}~\bibnamefont
  {Polli}}, \bibinfo {author} {\bibfnamefont {M.}~\bibnamefont {Rini}},
  \bibinfo {author} {\bibfnamefont {S.}~\bibnamefont {Wall}}, \bibinfo {author}
  {\bibfnamefont {R.~W.}\ \bibnamefont {Schoenlein}}, \bibinfo {author}
  {\bibfnamefont {Y.}~\bibnamefont {Tomioka}}, \bibinfo {author} {\bibfnamefont
  {Y.}~\bibnamefont {Tokura}}, \bibinfo {author} {\bibfnamefont
  {G.}~\bibnamefont {Cerullo}}, \ and\ \bibinfo {author} {\bibfnamefont
  {A.}~\bibnamefont {Cavalleri}},\ }\href {http://dx.doi.org/10.1038/nmat1979}
  {\bibfield  {journal} {\bibinfo  {journal} {Nat. Mater.}\ }\textbf {\bibinfo
  {volume} {6}},\ \bibinfo {pages} {643} (\bibinfo {year} {2007})}\BibitemShut
  {NoStop}%
\bibitem [{\citenamefont {Ehrke}\ \emph {et~al.}(2011)\citenamefont {Ehrke},
  \citenamefont {Tobey}, \citenamefont {Wall}, \citenamefont {Cavill},
  \citenamefont {F\"orst}, \citenamefont {Khanna}, \citenamefont {Garl},
  \citenamefont {Stojanovic}, \citenamefont {Prabhakaran}, \citenamefont
  {Boothroyd}, \citenamefont {Gensch}, \citenamefont {Mirone}, \citenamefont
  {Reutler}, \citenamefont {Revcolevschi}, \citenamefont {Dhesi},\ and\
  \citenamefont {Cavalleri}}]{Ehrke11}%
  \BibitemOpen
  \bibfield  {author} {\bibinfo {author} {\bibfnamefont {H.}~\bibnamefont
  {Ehrke}}, \bibinfo {author} {\bibfnamefont {R.~I.}\ \bibnamefont {Tobey}},
  \bibinfo {author} {\bibfnamefont {S.}~\bibnamefont {Wall}}, \bibinfo {author}
  {\bibfnamefont {S.~A.}\ \bibnamefont {Cavill}}, \bibinfo {author}
  {\bibfnamefont {M.}~\bibnamefont {F\"orst}}, \bibinfo {author} {\bibfnamefont
  {V.}~\bibnamefont {Khanna}}, \bibinfo {author} {\bibfnamefont
  {T.}~\bibnamefont {Garl}}, \bibinfo {author} {\bibfnamefont {N.}~\bibnamefont
  {Stojanovic}}, \bibinfo {author} {\bibfnamefont {D.}~\bibnamefont
  {Prabhakaran}}, \bibinfo {author} {\bibfnamefont {A.~T.}\ \bibnamefont
  {Boothroyd}}, \bibinfo {author} {\bibfnamefont {M.}~\bibnamefont {Gensch}},
  \bibinfo {author} {\bibfnamefont {A.}~\bibnamefont {Mirone}}, \bibinfo
  {author} {\bibfnamefont {P.}~\bibnamefont {Reutler}}, \bibinfo {author}
  {\bibfnamefont {A.}~\bibnamefont {Revcolevschi}}, \bibinfo {author}
  {\bibfnamefont {S.~S.}\ \bibnamefont {Dhesi}}, \ and\ \bibinfo {author}
  {\bibfnamefont {A.}~\bibnamefont {Cavalleri}},\ }\href {\doibase
  10.1103/PhysRevLett.106.217401} {\bibfield  {journal} {\bibinfo  {journal}
  {Phys. Rev. Lett.}\ }\textbf {\bibinfo {volume} {106}},\ \bibinfo {pages}
  {217401} (\bibinfo {year} {2011})}\BibitemShut {NoStop}%
\bibitem [{\citenamefont {Zhang}\ \emph {et~al.}(2016)\citenamefont {Zhang},
  \citenamefont {Tan}, \citenamefont {Liu}, \citenamefont {Teitelbaum},
  \citenamefont {Post}, \citenamefont {Jin}, \citenamefont {Nelson},
  \citenamefont {Basov}, \citenamefont {Wu},\ and\ \citenamefont
  {Averitt}}]{Zhang16}%
  \BibitemOpen
  \bibfield  {author} {\bibinfo {author} {\bibfnamefont {J.}~\bibnamefont
  {Zhang}}, \bibinfo {author} {\bibfnamefont {X.}~\bibnamefont {Tan}}, \bibinfo
  {author} {\bibfnamefont {M.}~\bibnamefont {Liu}}, \bibinfo {author}
  {\bibfnamefont {S.~W.}\ \bibnamefont {Teitelbaum}}, \bibinfo {author}
  {\bibfnamefont {K.~W.}\ \bibnamefont {Post}}, \bibinfo {author}
  {\bibfnamefont {F.}~\bibnamefont {Jin}}, \bibinfo {author} {\bibfnamefont
  {K.~A.}\ \bibnamefont {Nelson}}, \bibinfo {author} {\bibfnamefont {D.~N.}\
  \bibnamefont {Basov}}, \bibinfo {author} {\bibfnamefont {W.}~\bibnamefont
  {Wu}}, \ and\ \bibinfo {author} {\bibfnamefont {R.~D.}\ \bibnamefont
  {Averitt}},\ }\href {http://dx.doi.org/10.1038/nmat4695} {\bibfield
  {journal} {\bibinfo  {journal} {Nat. Mater.}\ }\textbf {\bibinfo {volume}
  {15}},\ \bibinfo {pages} {956} (\bibinfo {year} {2016})}\BibitemShut
  {NoStop}%
\bibitem [{\citenamefont {Casals}\ \emph {et~al.}(2016)\citenamefont {Casals},
  \citenamefont {Cichelero}, \citenamefont {Garc\'{\i}a~Fern\'andez},
  \citenamefont {Junquera}, \citenamefont {Pesquera}, \citenamefont
  {Campoy-Quiles}, \citenamefont {Infante}, \citenamefont {S\'anchez},
  \citenamefont {Fontcuberta},\ and\ \citenamefont {Herranz}}]{Casals16}%
  \BibitemOpen
  \bibfield  {author} {\bibinfo {author} {\bibfnamefont {B.}~\bibnamefont
  {Casals}}, \bibinfo {author} {\bibfnamefont {R.}~\bibnamefont {Cichelero}},
  \bibinfo {author} {\bibfnamefont {P.}~\bibnamefont
  {Garc\'{\i}a~Fern\'andez}}, \bibinfo {author} {\bibfnamefont
  {J.}~\bibnamefont {Junquera}}, \bibinfo {author} {\bibfnamefont
  {D.}~\bibnamefont {Pesquera}}, \bibinfo {author} {\bibfnamefont
  {M.}~\bibnamefont {Campoy-Quiles}}, \bibinfo {author} {\bibfnamefont {I.~C.}\
  \bibnamefont {Infante}}, \bibinfo {author} {\bibfnamefont {F.}~\bibnamefont
  {S\'anchez}}, \bibinfo {author} {\bibfnamefont {J.}~\bibnamefont
  {Fontcuberta}}, \ and\ \bibinfo {author} {\bibfnamefont {G.}~\bibnamefont
  {Herranz}},\ }\href {\doibase 10.1103/PhysRevLett.117.026401} {\bibfield
  {journal} {\bibinfo  {journal} {Phys. Rev. Lett.}\ }\textbf {\bibinfo
  {volume} {117}},\ \bibinfo {pages} {026401} (\bibinfo {year}
  {2016})}\BibitemShut {NoStop}%
\bibitem [{\citenamefont {Fox}(2010)}]{Fox2010}%
  \BibitemOpen
  \bibfield  {author} {\bibinfo {author} {\bibfnamefont {M.}~\bibnamefont
  {Fox}},\ }\href@noop {} {\emph {\bibinfo {title} {Optical Properties of
  Solids (Oxford Master Series in Condensed Matter Physics)}}},\ \bibinfo
  {edition} {2nd}\ ed.\ (\bibinfo  {publisher} {Oxford University Press, USA},\
  \bibinfo {year} {2010})\BibitemShut {NoStop}%
\bibitem [{\citenamefont {Wooten}(2013)}]{Wooten2013}%
  \BibitemOpen
  \bibfield  {author} {\bibinfo {author} {\bibfnamefont {F.}~\bibnamefont
  {Wooten}},\ }\href@noop {} {\emph {\bibinfo {title} {Optical properties of
  solids}}}\ (\bibinfo  {publisher} {Academic press, New York},\ \bibinfo
  {year} {2013})\BibitemShut {NoStop}%
\bibitem [{\citenamefont {Mitrano}\ \emph {et~al.}(2014)\citenamefont
  {Mitrano}, \citenamefont {Cotugno}, \citenamefont {Clark}, \citenamefont
  {Singla}, \citenamefont {Kaiser}, \citenamefont {St\"ahler}, \citenamefont
  {Beyer}, \citenamefont {Dressel}, \citenamefont {Baldassarre}, \citenamefont
  {Nicoletti}, \citenamefont {Perucchi}, \citenamefont {Hasegawa},
  \citenamefont {Okamoto}, \citenamefont {Jaksch},\ and\ \citenamefont
  {Cavalleri}}]{Mitrano14}%
  \BibitemOpen
  \bibfield  {author} {\bibinfo {author} {\bibfnamefont {M.}~\bibnamefont
  {Mitrano}}, \bibinfo {author} {\bibfnamefont {G.}~\bibnamefont {Cotugno}},
  \bibinfo {author} {\bibfnamefont {S.~R.}\ \bibnamefont {Clark}}, \bibinfo
  {author} {\bibfnamefont {R.}~\bibnamefont {Singla}}, \bibinfo {author}
  {\bibfnamefont {S.}~\bibnamefont {Kaiser}}, \bibinfo {author} {\bibfnamefont
  {J.}~\bibnamefont {St\"ahler}}, \bibinfo {author} {\bibfnamefont
  {R.}~\bibnamefont {Beyer}}, \bibinfo {author} {\bibfnamefont
  {M.}~\bibnamefont {Dressel}}, \bibinfo {author} {\bibfnamefont
  {L.}~\bibnamefont {Baldassarre}}, \bibinfo {author} {\bibfnamefont
  {D.}~\bibnamefont {Nicoletti}}, \bibinfo {author} {\bibfnamefont
  {A.}~\bibnamefont {Perucchi}}, \bibinfo {author} {\bibfnamefont
  {T.}~\bibnamefont {Hasegawa}}, \bibinfo {author} {\bibfnamefont
  {H.}~\bibnamefont {Okamoto}}, \bibinfo {author} {\bibfnamefont
  {D.}~\bibnamefont {Jaksch}}, \ and\ \bibinfo {author} {\bibfnamefont
  {A.}~\bibnamefont {Cavalleri}},\ }\href {\doibase
  10.1103/PhysRevLett.112.117801} {\bibfield  {journal} {\bibinfo  {journal}
  {Phys. Rev. Lett.}\ }\textbf {\bibinfo {volume} {112}},\ \bibinfo {pages}
  {117801} (\bibinfo {year} {2014})}\BibitemShut {NoStop}%
\bibitem [{\citenamefont {Oka}\ \emph {et~al.}(2003)\citenamefont {Oka},
  \citenamefont {Arita},\ and\ \citenamefont {Aoki}}]{Oka03}%
  \BibitemOpen
  \bibfield  {author} {\bibinfo {author} {\bibfnamefont {T.}~\bibnamefont
  {Oka}}, \bibinfo {author} {\bibfnamefont {R.}~\bibnamefont {Arita}}, \ and\
  \bibinfo {author} {\bibfnamefont {H.}~\bibnamefont {Aoki}},\ }\href {\doibase
  10.1103/PhysRevLett.91.066406} {\bibfield  {journal} {\bibinfo  {journal}
  {Phys. Rev. Lett.}\ }\textbf {\bibinfo {volume} {91}},\ \bibinfo {pages}
  {066406} (\bibinfo {year} {2003})}\BibitemShut {NoStop}%
\bibitem [{\citenamefont {Oka}(2012)}]{Oka12}%
  \BibitemOpen
  \bibfield  {author} {\bibinfo {author} {\bibfnamefont {T.}~\bibnamefont
  {Oka}},\ }\href {\doibase 10.1103/PhysRevB.86.075148} {\bibfield  {journal}
  {\bibinfo  {journal} {Phys. Rev. B}\ }\textbf {\bibinfo {volume} {86}},\
  \bibinfo {pages} {075148} (\bibinfo {year} {2012})}\BibitemShut {NoStop}%
\bibitem [{\citenamefont {Al-Hassanieh}\ \emph {et~al.}(2008)\citenamefont
  {Al-Hassanieh}, \citenamefont {Reboredo}, \citenamefont {Feiguin},
  \citenamefont {Gonz\'alez},\ and\ \citenamefont {Dagotto}}]{Khaled08}%
  \BibitemOpen
  \bibfield  {author} {\bibinfo {author} {\bibfnamefont {K.~A.}\ \bibnamefont
  {Al-Hassanieh}}, \bibinfo {author} {\bibfnamefont {F.~A.}\ \bibnamefont
  {Reboredo}}, \bibinfo {author} {\bibfnamefont {A.~E.}\ \bibnamefont
  {Feiguin}}, \bibinfo {author} {\bibfnamefont {I.}~\bibnamefont {Gonz\'alez}},
  \ and\ \bibinfo {author} {\bibfnamefont {E.}~\bibnamefont {Dagotto}},\ }\href
  {\doibase 10.1103/PhysRevLett.100.166403} {\bibfield  {journal} {\bibinfo
  {journal} {Phys. Rev. Lett.}\ }\textbf {\bibinfo {volume} {100}},\ \bibinfo
  {pages} {166403} (\bibinfo {year} {2008})}\BibitemShut {NoStop}%
\bibitem [{\citenamefont {Heidrich-Meisner}\ \emph {et~al.}(2010)\citenamefont
  {Heidrich-Meisner}, \citenamefont {Gonz\'alez}, \citenamefont {Al-Hassanieh},
  \citenamefont {Feiguin}, \citenamefont {Rozenberg},\ and\ \citenamefont
  {Dagotto}}]{Fabian10}%
  \BibitemOpen
  \bibfield  {author} {\bibinfo {author} {\bibfnamefont {F.}~\bibnamefont
  {Heidrich-Meisner}}, \bibinfo {author} {\bibfnamefont {I.}~\bibnamefont
  {Gonz\'alez}}, \bibinfo {author} {\bibfnamefont {K.~A.}\ \bibnamefont
  {Al-Hassanieh}}, \bibinfo {author} {\bibfnamefont {A.~E.}\ \bibnamefont
  {Feiguin}}, \bibinfo {author} {\bibfnamefont {M.~J.}\ \bibnamefont
  {Rozenberg}}, \ and\ \bibinfo {author} {\bibfnamefont {E.}~\bibnamefont
  {Dagotto}},\ }\href {\doibase 10.1103/PhysRevB.82.205110} {\bibfield
  {journal} {\bibinfo  {journal} {Phys. Rev. B}\ }\textbf {\bibinfo {volume}
  {82}},\ \bibinfo {pages} {205110} (\bibinfo {year} {2010})}\BibitemShut
  {NoStop}%
\bibitem [{\citenamefont {Hofmann}\ and\ \citenamefont
  {Potthoff}(2012)}]{Hofmann12}%
  \BibitemOpen
  \bibfield  {author} {\bibinfo {author} {\bibfnamefont {F.}~\bibnamefont
  {Hofmann}}\ and\ \bibinfo {author} {\bibfnamefont {M.}~\bibnamefont
  {Potthoff}},\ }\href {\doibase 10.1103/PhysRevB.85.205127} {\bibfield
  {journal} {\bibinfo  {journal} {Phys. Rev. B}\ }\textbf {\bibinfo {volume}
  {85}},\ \bibinfo {pages} {205127} (\bibinfo {year} {2012})}\BibitemShut
  {NoStop}%
\bibitem [{\citenamefont {Eckstein}\ \emph {et~al.}(2010)\citenamefont
  {Eckstein}, \citenamefont {Oka},\ and\ \citenamefont {Werner}}]{Eck10}%
  \BibitemOpen
  \bibfield  {author} {\bibinfo {author} {\bibfnamefont {M.}~\bibnamefont
  {Eckstein}}, \bibinfo {author} {\bibfnamefont {T.}~\bibnamefont {Oka}}, \
  and\ \bibinfo {author} {\bibfnamefont {P.}~\bibnamefont {Werner}},\ }\href
  {\doibase 10.1103/PhysRevLett.105.146404} {\bibfield  {journal} {\bibinfo
  {journal} {Phys. Rev. Lett.}\ }\textbf {\bibinfo {volume} {105}},\ \bibinfo
  {pages} {146404} (\bibinfo {year} {2010})}\BibitemShut {NoStop}%
\bibitem [{\citenamefont {Eckstein}\ and\ \citenamefont
  {Werner}(2013)}]{Eck13}%
  \BibitemOpen
  \bibfield  {author} {\bibinfo {author} {\bibfnamefont {M.}~\bibnamefont
  {Eckstein}}\ and\ \bibinfo {author} {\bibfnamefont {P.}~\bibnamefont
  {Werner}},\ }\href {\doibase 10.1103/PhysRevLett.110.126401} {\bibfield
  {journal} {\bibinfo  {journal} {Phys. Rev. Lett.}\ }\textbf {\bibinfo
  {volume} {110}},\ \bibinfo {pages} {126401} (\bibinfo {year}
  {2013})}\BibitemShut {NoStop}%
\bibitem [{\citenamefont {Kanamori}\ \emph {et~al.}(2009)\citenamefont
  {Kanamori}, \citenamefont {Matsueda},\ and\ \citenamefont
  {Ishihara}}]{Kanamori09}%
  \BibitemOpen
  \bibfield  {author} {\bibinfo {author} {\bibfnamefont {Y.}~\bibnamefont
  {Kanamori}}, \bibinfo {author} {\bibfnamefont {H.}~\bibnamefont {Matsueda}},
  \ and\ \bibinfo {author} {\bibfnamefont {S.}~\bibnamefont {Ishihara}},\
  }\href {\doibase 10.1103/PhysRevLett.103.267401} {\bibfield  {journal}
  {\bibinfo  {journal} {Phys. Rev. Lett.}\ }\textbf {\bibinfo {volume} {103}},\
  \bibinfo {pages} {267401} (\bibinfo {year} {2009})}\BibitemShut {NoStop}%
\bibitem [{\citenamefont {Maeshima}\ \emph {et~al.}(2010)\citenamefont
  {Maeshima}, \citenamefont {Hino},\ and\ \citenamefont
  {Yonemitsu}}]{Maeshima10prb}%
  \BibitemOpen
  \bibfield  {author} {\bibinfo {author} {\bibfnamefont {N.}~\bibnamefont
  {Maeshima}}, \bibinfo {author} {\bibfnamefont {K.}~\bibnamefont {Hino}}, \
  and\ \bibinfo {author} {\bibfnamefont {K.}~\bibnamefont {Yonemitsu}},\ }\href
  {\doibase 10.1103/PhysRevB.82.161105} {\bibfield  {journal} {\bibinfo
  {journal} {Phys. Rev. B}\ }\textbf {\bibinfo {volume} {82}},\ \bibinfo
  {pages} {161105} (\bibinfo {year} {2010})}\BibitemShut {NoStop}%
\bibitem [{\citenamefont {Kanamori}\ \emph {et~al.}(2011)\citenamefont
  {Kanamori}, \citenamefont {Matsueda},\ and\ \citenamefont
  {Ishihara}}]{Kanamori11}%
  \BibitemOpen
  \bibfield  {author} {\bibinfo {author} {\bibfnamefont {Y.}~\bibnamefont
  {Kanamori}}, \bibinfo {author} {\bibfnamefont {H.}~\bibnamefont {Matsueda}},
  \ and\ \bibinfo {author} {\bibfnamefont {S.}~\bibnamefont {Ishihara}},\
  }\href {\doibase 10.1103/PhysRevLett.107.167403} {\bibfield  {journal}
  {\bibinfo  {journal} {Phys. Rev. Lett.}\ }\textbf {\bibinfo {volume} {107}},\
  \bibinfo {pages} {167403} (\bibinfo {year} {2011})}\BibitemShut {NoStop}%
\bibitem [{\citenamefont {Kanamori}\ \emph {et~al.}(2012)\citenamefont
  {Kanamori}, \citenamefont {Ohara},\ and\ \citenamefont
  {Ishihara}}]{Kanamori12}%
  \BibitemOpen
  \bibfield  {author} {\bibinfo {author} {\bibfnamefont {Y.}~\bibnamefont
  {Kanamori}}, \bibinfo {author} {\bibfnamefont {J.}~\bibnamefont {Ohara}}, \
  and\ \bibinfo {author} {\bibfnamefont {S.}~\bibnamefont {Ishihara}},\ }\href
  {\doibase 10.1103/PhysRevB.86.045137} {\bibfield  {journal} {\bibinfo
  {journal} {Phys. Rev. B}\ }\textbf {\bibinfo {volume} {86}},\ \bibinfo
  {pages} {045137} (\bibinfo {year} {2012})}\BibitemShut {NoStop}%
\bibitem [{\citenamefont {Ishihara}\ \emph {et~al.}(2013)\citenamefont
  {Ishihara}, \citenamefont {Ohara},\ and\ \citenamefont
  {Kanamori}}]{Ishihara13}%
  \BibitemOpen
  \bibfield  {author} {\bibinfo {author} {\bibfnamefont {S.}~\bibnamefont
  {Ishihara}}, \bibinfo {author} {\bibfnamefont {J.}~\bibnamefont {Ohara}}, \
  and\ \bibinfo {author} {\bibfnamefont {Y.}~\bibnamefont {Kanamori}},\ }\href
  {\doibase 10.1140/epjst/e2013-01910-4} {\bibfield  {journal} {\bibinfo
  {journal} {Eur. Phys. J. Spec. Top.}\ }\textbf {\bibinfo {volume} {222}},\
  \bibinfo {pages} {1125} (\bibinfo {year} {2013})}\BibitemShut {NoStop}%
\bibitem [{\citenamefont {White}(1992)}]{dmrg1}%
  \BibitemOpen
  \bibfield  {author} {\bibinfo {author} {\bibfnamefont {S.~R.}\ \bibnamefont
  {White}},\ }\href {\doibase 10.1103/PhysRevLett.69.2863} {\bibfield
  {journal} {\bibinfo  {journal} {Phys. Rev. Lett.}\ }\textbf {\bibinfo
  {volume} {69}},\ \bibinfo {pages} {2863} (\bibinfo {year}
  {1992})}\BibitemShut {NoStop}%
\bibitem [{\citenamefont {White}(1993)}]{dmrg2}%
  \BibitemOpen
  \bibfield  {author} {\bibinfo {author} {\bibfnamefont {S.~R.}\ \bibnamefont
  {White}},\ }\href {\doibase 10.1103/PhysRevB.48.10345} {\bibfield  {journal}
  {\bibinfo  {journal} {Phys. Rev. B}\ }\textbf {\bibinfo {volume} {48}},\
  \bibinfo {pages} {10345} (\bibinfo {year} {1993})}\BibitemShut {NoStop}%
\bibitem [{\citenamefont {Schollw\"ock}(2005)}]{dmrg3}%
  \BibitemOpen
  \bibfield  {author} {\bibinfo {author} {\bibfnamefont {U.}~\bibnamefont
  {Schollw\"ock}},\ }\href {\doibase 10.1103/RevModPhys.77.259} {\bibfield
  {journal} {\bibinfo  {journal} {Rev. Mod. Phys.}\ }\textbf {\bibinfo {volume}
  {77}},\ \bibinfo {pages} {259} (\bibinfo {year} {2005})}\BibitemShut
  {NoStop}%
\bibitem [{\citenamefont {Hallberg}(2006)}]{dmrg4}%
  \BibitemOpen
  \bibfield  {author} {\bibinfo {author} {\bibfnamefont {K.~A.}\ \bibnamefont
  {Hallberg}},\ }\href {\doibase 10.1080/00018730600766432} {\bibfield
  {journal} {\bibinfo  {journal} {Adv. Phys.}\ }\textbf {\bibinfo {volume}
  {55}},\ \bibinfo {pages} {477} (\bibinfo {year} {2006})}\BibitemShut
  {NoStop}%
\bibitem [{\citenamefont {White}\ and\ \citenamefont
  {Feiguin}(2004)}]{White04}%
  \BibitemOpen
  \bibfield  {author} {\bibinfo {author} {\bibfnamefont {S.~R.}\ \bibnamefont
  {White}}\ and\ \bibinfo {author} {\bibfnamefont {A.~E.}\ \bibnamefont
  {Feiguin}},\ }\href {\doibase 10.1103/PhysRevLett.93.076401} {\bibfield
  {journal} {\bibinfo  {journal} {Phys. Rev. Lett.}\ }\textbf {\bibinfo
  {volume} {93}},\ \bibinfo {pages} {076401} (\bibinfo {year}
  {2004})}\BibitemShut {NoStop}%
\bibitem [{\citenamefont {Daley}\ \emph {et~al.}(2004)\citenamefont {Daley},
  \citenamefont {Kollath}, \citenamefont {Schollw\"ock},\ and\ \citenamefont
  {Vidal}}]{Daley04}%
  \BibitemOpen
  \bibfield  {author} {\bibinfo {author} {\bibfnamefont {A.~J.}\ \bibnamefont
  {Daley}}, \bibinfo {author} {\bibfnamefont {C.}~\bibnamefont {Kollath}},
  \bibinfo {author} {\bibfnamefont {U.}~\bibnamefont {Schollw\"ock}}, \ and\
  \bibinfo {author} {\bibfnamefont {G.}~\bibnamefont {Vidal}},\ }\href
  {http://stacks.iop.org/1742-5468/2004/i=04/a=P04005} {\bibfield  {journal}
  {\bibinfo  {journal} {J. Stat. Mech.: Theo. Exp.}\ }\textbf {\bibinfo
  {volume} {2004}},\ \bibinfo {pages} {P04005} (\bibinfo {year}
  {2004})}\BibitemShut {NoStop}%
\bibitem [{\citenamefont {Feiguin}\ and\ \citenamefont
  {White}(2005)}]{Feiguin05}%
  \BibitemOpen
  \bibfield  {author} {\bibinfo {author} {\bibfnamefont {A.~E.}\ \bibnamefont
  {Feiguin}}\ and\ \bibinfo {author} {\bibfnamefont {S.~R.}\ \bibnamefont
  {White}},\ }\href {\doibase 10.1103/PhysRevB.72.020404} {\bibfield  {journal}
  {\bibinfo  {journal} {Phys. Rev. B}\ }\textbf {\bibinfo {volume} {72}},\
  \bibinfo {pages} {020404} (\bibinfo {year} {2005})}\BibitemShut {NoStop}%
\bibitem [{\citenamefont {Schmitteckert}(2004)}]{rk4}%
  \BibitemOpen
  \bibfield  {author} {\bibinfo {author} {\bibfnamefont {P.}~\bibnamefont
  {Schmitteckert}},\ }\href {\doibase 10.1103/PhysRevB.70.121302} {\bibfield
  {journal} {\bibinfo  {journal} {Phys. Rev. B}\ }\textbf {\bibinfo {volume}
  {70}},\ \bibinfo {pages} {121302} (\bibinfo {year} {2004})}\BibitemShut
  {NoStop}%
\bibitem [{\citenamefont {Rinc\'on}\ \emph
  {et~al.}(2014{\natexlab{a}})\citenamefont {Rinc\'on}, \citenamefont {Moreo},
  \citenamefont {Alvarez},\ and\ \citenamefont {Dagotto}}]{Rincon14prl}%
  \BibitemOpen
  \bibfield  {author} {\bibinfo {author} {\bibfnamefont {J.}~\bibnamefont
  {Rinc\'on}}, \bibinfo {author} {\bibfnamefont {A.}~\bibnamefont {Moreo}},
  \bibinfo {author} {\bibfnamefont {G.}~\bibnamefont {Alvarez}}, \ and\
  \bibinfo {author} {\bibfnamefont {E.}~\bibnamefont {Dagotto}},\ }\href
  {\doibase 10.1103/PhysRevLett.112.106405} {\bibfield  {journal} {\bibinfo
  {journal} {Phys. Rev. Lett.}\ }\textbf {\bibinfo {volume} {112}},\ \bibinfo
  {pages} {106405} (\bibinfo {year} {2014}{\natexlab{a}})}\BibitemShut
  {NoStop}%
\bibitem [{\citenamefont {Rinc\'on}\ \emph
  {et~al.}(2014{\natexlab{b}})\citenamefont {Rinc\'on}, \citenamefont {Moreo},
  \citenamefont {Alvarez},\ and\ \citenamefont {Dagotto}}]{Rincon14}%
  \BibitemOpen
  \bibfield  {author} {\bibinfo {author} {\bibfnamefont {J.}~\bibnamefont
  {Rinc\'on}}, \bibinfo {author} {\bibfnamefont {A.}~\bibnamefont {Moreo}},
  \bibinfo {author} {\bibfnamefont {G.}~\bibnamefont {Alvarez}}, \ and\
  \bibinfo {author} {\bibfnamefont {E.}~\bibnamefont {Dagotto}},\ }\href
  {\doibase 10.1103/PhysRevB.90.241105} {\bibfield  {journal} {\bibinfo
  {journal} {Phys. Rev. B}\ }\textbf {\bibinfo {volume} {90}},\ \bibinfo
  {pages} {241105} (\bibinfo {year} {2014}{\natexlab{b}})}\BibitemShut
  {NoStop}%
\bibitem [{\citenamefont {Dai}\ \emph {et~al.}(2012)\citenamefont {Dai},
  \citenamefont {Hu},\ and\ \citenamefont {Dagotto}}]{Dagotto12}%
  \BibitemOpen
  \bibfield  {author} {\bibinfo {author} {\bibfnamefont {P.}~\bibnamefont
  {Dai}}, \bibinfo {author} {\bibfnamefont {J.}~\bibnamefont {Hu}}, \ and\
  \bibinfo {author} {\bibfnamefont {E.}~\bibnamefont {Dagotto}},\ }\href
  {http://dx.doi.org/10.1038/nphys2438} {\bibfield  {journal} {\bibinfo
  {journal} {Nat. Phys.}\ }\textbf {\bibinfo {volume} {8}},\ \bibinfo {pages}
  {709} (\bibinfo {year} {2012})}\BibitemShut {NoStop}%
\bibitem [{\citenamefont {Georges}\ \emph {et~al.}(2013)\citenamefont
  {Georges}, \citenamefont {de' Medici},\ and\ \citenamefont
  {Mravlje}}]{Georges13}%
  \BibitemOpen
  \bibfield  {author} {\bibinfo {author} {\bibfnamefont {A.}~\bibnamefont
  {Georges}}, \bibinfo {author} {\bibfnamefont {L.}~\bibnamefont {de' Medici}},
  \ and\ \bibinfo {author} {\bibfnamefont {J.}~\bibnamefont {Mravlje}},\ }\href
  {\doibase 10.1146/annurev-conmatphys-020911-125045} {\bibfield  {journal}
  {\bibinfo  {journal} {Annu. Rev. Condens. Matter Phys.}\ }\textbf {\bibinfo
  {volume} {4}},\ \bibinfo {pages} {137} (\bibinfo {year} {2013})}\BibitemShut
  {NoStop}%
\bibitem [{\citenamefont {Dagotto}(1994)}]{Dagotto94}%
  \BibitemOpen
  \bibfield  {author} {\bibinfo {author} {\bibfnamefont {E.}~\bibnamefont
  {Dagotto}},\ }\href {\doibase 10.1103/RevModPhys.66.763} {\bibfield
  {journal} {\bibinfo  {journal} {Rev. Mod. Phys.}\ }\textbf {\bibinfo {volume}
  {66}},\ \bibinfo {pages} {763} (\bibinfo {year} {1994})}\BibitemShut
  {NoStop}%
\bibitem [{\citenamefont {Al-Hassanieh}\ \emph {et~al.}(2009)\citenamefont
  {Al-Hassanieh}, \citenamefont {Batista}, \citenamefont {Sengupta},\ and\
  \citenamefont {Feiguin}}]{Khaled09}%
  \BibitemOpen
  \bibfield  {author} {\bibinfo {author} {\bibfnamefont {K.~A.}\ \bibnamefont
  {Al-Hassanieh}}, \bibinfo {author} {\bibfnamefont {C.~D.}\ \bibnamefont
  {Batista}}, \bibinfo {author} {\bibfnamefont {P.}~\bibnamefont {Sengupta}}, \
  and\ \bibinfo {author} {\bibfnamefont {A.~E.}\ \bibnamefont {Feiguin}},\
  }\href {\doibase 10.1103/PhysRevB.80.115116} {\bibfield  {journal} {\bibinfo
  {journal} {Phys. Rev. B}\ }\textbf {\bibinfo {volume} {80}},\ \bibinfo
  {pages} {115116} (\bibinfo {year} {2009})}\BibitemShut {NoStop}%
\bibitem [{\citenamefont {Grenier}\ \emph {et~al.}(2015)\citenamefont
  {Grenier}, \citenamefont {Petit}, \citenamefont {Simonet}, \citenamefont
  {Can\'evet}, \citenamefont {Regnault}, \citenamefont {Raymond}, \citenamefont
  {Canals}, \citenamefont {Berthier},\ and\ \citenamefont {Lejay}}]{Grenier15}%
  \BibitemOpen
  \bibfield  {author} {\bibinfo {author} {\bibfnamefont {B.}~\bibnamefont
  {Grenier}}, \bibinfo {author} {\bibfnamefont {S.}~\bibnamefont {Petit}},
  \bibinfo {author} {\bibfnamefont {V.}~\bibnamefont {Simonet}}, \bibinfo
  {author} {\bibfnamefont {E.}~\bibnamefont {Can\'evet}}, \bibinfo {author}
  {\bibfnamefont {L.-P.}\ \bibnamefont {Regnault}}, \bibinfo {author}
  {\bibfnamefont {S.}~\bibnamefont {Raymond}}, \bibinfo {author} {\bibfnamefont
  {B.}~\bibnamefont {Canals}}, \bibinfo {author} {\bibfnamefont
  {C.}~\bibnamefont {Berthier}}, \ and\ \bibinfo {author} {\bibfnamefont
  {P.}~\bibnamefont {Lejay}},\ }\href {\doibase 10.1103/PhysRevLett.114.017201}
  {\bibfield  {journal} {\bibinfo  {journal} {Phys. Rev. Lett.}\ }\textbf
  {\bibinfo {volume} {114}},\ \bibinfo {pages} {017201} (\bibinfo {year}
  {2015})}\BibitemShut {NoStop}%
\bibitem [{\citenamefont {Bera}\ \emph {et~al.}(2017)\citenamefont {Bera},
  \citenamefont {Lake}, \citenamefont {Essler}, \citenamefont {Vanderstraeten},
  \citenamefont {Hubig}, \citenamefont {Schollw\"ock}, \citenamefont {Islam},
  \citenamefont {Schneidewind},\ and\ \citenamefont
  {Quintero-Castro}}]{Bera17}%
  \BibitemOpen
  \bibfield  {author} {\bibinfo {author} {\bibfnamefont {A.~K.}\ \bibnamefont
  {Bera}}, \bibinfo {author} {\bibfnamefont {B.}~\bibnamefont {Lake}}, \bibinfo
  {author} {\bibfnamefont {F.~H.~L.}\ \bibnamefont {Essler}}, \bibinfo {author}
  {\bibfnamefont {L.}~\bibnamefont {Vanderstraeten}}, \bibinfo {author}
  {\bibfnamefont {C.}~\bibnamefont {Hubig}}, \bibinfo {author} {\bibfnamefont
  {U.}~\bibnamefont {Schollw\"ock}}, \bibinfo {author} {\bibfnamefont {A.~T.
  M.~N.}\ \bibnamefont {Islam}}, \bibinfo {author} {\bibfnamefont
  {A.}~\bibnamefont {Schneidewind}}, \ and\ \bibinfo {author} {\bibfnamefont
  {D.~L.}\ \bibnamefont {Quintero-Castro}},\ }\href {\doibase
  10.1103/PhysRevB.96.054423} {\bibfield  {journal} {\bibinfo  {journal} {Phys.
  Rev. B}\ }\textbf {\bibinfo {volume} {96}},\ \bibinfo {pages} {054423}
  (\bibinfo {year} {2017})}\BibitemShut {NoStop}%
\bibitem [{\citenamefont {Faure}\ \emph {et~al.}(2017)\citenamefont {Faure},
  \citenamefont {Takayoshi}, \citenamefont {Petit}, \citenamefont {Simonet},
  \citenamefont {Raymond}, \citenamefont {Regnault}, \citenamefont {Boehm},
  \citenamefont {White}, \citenamefont {M{\aa}nsson}, \citenamefont
  {R{\"u}egg}, \citenamefont {Lejay}, \citenamefont {Canals}, \citenamefont
  {Furuya}, \citenamefont {Giamarchi},\ and\ \citenamefont
  {Grenier}}]{Faure17}%
  \BibitemOpen
  \bibfield  {author} {\bibinfo {author} {\bibfnamefont {Q.}~\bibnamefont
  {Faure}}, \bibinfo {author} {\bibfnamefont {S.}~\bibnamefont {Takayoshi}},
  \bibinfo {author} {\bibfnamefont {S.}~\bibnamefont {Petit}}, \bibinfo
  {author} {\bibfnamefont {V.}~\bibnamefont {Simonet}}, \bibinfo {author}
  {\bibfnamefont {S.}~\bibnamefont {Raymond}}, \bibinfo {author} {\bibfnamefont
  {L.-P.}\ \bibnamefont {Regnault}}, \bibinfo {author} {\bibfnamefont
  {M.}~\bibnamefont {Boehm}}, \bibinfo {author} {\bibfnamefont
  {J.}~\bibnamefont {White}}, \bibinfo {author} {\bibfnamefont
  {M.}~\bibnamefont {M{\aa}nsson}}, \bibinfo {author} {\bibfnamefont
  {C.}~\bibnamefont {R{\"u}egg}}, \bibinfo {author} {\bibfnamefont
  {P.}~\bibnamefont {Lejay}}, \bibinfo {author} {\bibfnamefont
  {B.}~\bibnamefont {Canals}}, \bibinfo {author} {\bibfnamefont
  {S.}~\bibnamefont {Furuya}}, \bibinfo {author} {\bibfnamefont
  {T.}~\bibnamefont {Giamarchi}}, \ and\ \bibinfo {author} {\bibfnamefont
  {B.}~\bibnamefont {Grenier}},\ }\href {https://arxiv.org/abs/1706.05848}
  {\bibfield  {journal} {\bibinfo  {journal} {arXiv:1706.05848}\ } (\bibinfo
  {year} {2017})}\BibitemShut {NoStop}%
\bibitem [{\citenamefont {Strand}\ \emph {et~al.}(2017)\citenamefont {Strand},
  \citenamefont {Gole\ifmmode~\check{z}\else \v{z}\fi{}}, \citenamefont
  {Eckstein},\ and\ \citenamefont {Werner}}]{Strand17}%
  \BibitemOpen
  \bibfield  {author} {\bibinfo {author} {\bibfnamefont {H.~U.~R.}\
  \bibnamefont {Strand}}, \bibinfo {author} {\bibfnamefont {D.}~\bibnamefont
  {Gole\ifmmode~\check{z}\else \v{z}\fi{}}}, \bibinfo {author} {\bibfnamefont
  {M.}~\bibnamefont {Eckstein}}, \ and\ \bibinfo {author} {\bibfnamefont
  {P.}~\bibnamefont {Werner}},\ }\href {\doibase 10.1103/PhysRevB.96.165104}
  {\bibfield  {journal} {\bibinfo  {journal} {Phys. Rev. B}\ }\textbf {\bibinfo
  {volume} {96}},\ \bibinfo {pages} {165104} (\bibinfo {year}
  {2017})}\BibitemShut {NoStop}%
\end{thebibliography}%

\end{document}